\documentclass[useAMS]{mn2e}
\usepackage{times,graphicx}
\def\DS     {\displaystyle}
\def\E#1{\hbox{$10^{#1}$}}
\def\eq#1{\begin{equation} #1 \end{equation}}
\def\Figure#1#2
           {\centering\leavevmode\includegraphics[width=#2,clip]{#1.ps}}
\def\eqarray#1{\begin{eqnarray} #1 \end{eqnarray}}
\def\non{\nonumber \\}
\def\DnuD     {\hbox{$\Delta\nu_D$}}
\def\Jbar     {\hbox{$\bar J$}}
\def\j        {\hbox{$\jmath$}}
\def\N        {\hbox{$\cal N$}}
\def\Ie       {\hbox{$I_e$}}
\def\Ji       {\hbox{$\bar J^i_e$}}
\def\about    {\hbox{$\sim$}}
\def\x        {\hbox{$\times$}}
\def\half     {\hbox{$1\over2$}}
\def\Ncr      {\hbox{$N'_{\rm cr}$}}
\def\mic      {\hbox{$\mu$m}}
\def\ion#1#2  {#1\,{\small {#2}} }
\def\tot      {\tau_t}
\def\t(#1){\tau^{#1}}
\def\a(#1){\alpha^{#1}}

\newcount\hour
\newcount\minute
\def\clock    {\hour = \time \divide\hour by 60 \minute = \hour
                 \multiply\minute by -60  \advance\minute by \time
                 \number\hour:\ifnum\minute < 10 {0\number\minute}
                                              \else\number\minute
                                              \fi}
\def\Draft{Submitted August 8, 2005; accepted October 20, 2005}

\title[Exact Line Transfer]
{A New Exact Method for Line Radiative Transfer}

\author[Elitzur \& Asensio Ramos]
   {Moshe Elitzur$^1$ and Andr\'es Asensio Ramos$^2$\\
$^1$Department of Physics \& Astronomy, University of Kentucky,
    Lexington, KY 40506, USA; moshe@pa.uky.edu \\
$^2$INAF-Osservatorio Astrofisico di Arcetri, Largo E. Fermi 5,
    50125, Firenze, Italy; aasensio@arcetri.astro.it
}

\date{\Draft}
\pagerange{\pageref{firstpage}--\pageref{lastpage}} \pubyear{2005}

\begin{document}
\label{firstpage}

\maketitle

\begin{abstract}
We present a new method, the Coupled Escape Probability (CEP), for exact
calculation of line emission from multi-level systems, solving only algebraic
equations for the level populations. The CEP formulation of the classical
two-level problem is a set of {\em linear equations}, and we uncover an exact
analytic expression for the emission from two-level optically thick sources
that holds as long as they are in the ``effectively thin" regime. In
comparative study of a number of standard problems, the CEP method outperformed
the leading line transfer methods by substantial margins.

The algebraic equations employed by our new method are already incorporated in
numerous codes based on the escape probability approximation. All that is
required for an exact solution with these existing codes is to augment the
expression for the escape probability with simple zone-coupling terms. As an
application, we find that standard escape probability calculations generally
produce the correct cooling emission by the \ion{C}{II} 158 \mic\ line but not
by the $^3$P lines of \ion{O}{I}.

\end{abstract}


\section{INTRODUCTION}

Much of the information about astronomical sources comes from spectral lines,
requiring reliable analysis of multi-level line emission. Most current methods
for exact solutions involve accelerated $\Lambda$-iteration (ALI)
techniques\footnote{As noted by Trujillo Bueno \& Fabiani Bendicho (1995), the
ALI method is based on the Jacobi iteration (Jacobi 1845).} in which the
radiation intensity is obtained from the repeated action of an operator
designated $\Lambda$ on the source function (e.g., Rybicki 1991, Hubeny 1992).
The ALI method utilizing short characteristics with parabolic interpolation of
the source function (hereafter SCP; Olson, Auer \& Buchler 1986) is a standard
against which the efficiency of other line transfer techniques can be measured.

Because of the complexity and computational demands of exact methods, many
simulation codes that attempt to implement as many realistic physical
ingredients as possible are altogether bypassing solution of the radiative
transfer equation, employing instead the escape probability technique. In this
approach only the level populations are considered, calculated from rate
equations that include photon escape factors which are meant to account
approximately for the effects of radiative transfer (see Dumont et al 2003 for
a recent discussion and comparison with ALI calculations). Since this approach
is founded on a plausibility assumption right from the start, its results
amount to an uncontrolled approximation without any means for internal error
estimates. Nevertheless, this inherent shortcoming is often tolerated because
of the simplicity and usefulness of the escape probability approach.

We present here a new exact method, the Coupled Escape Probability (CEP), that
retains all the advantages of the naive escape probability approach. In this
new technique the source is divided into zones, and formal level population
equations that are fully consistent with radiative transfer are derived
rigorously from first principles. Different zones are coupled through terms
resembling standard escape probability expressions, resulting in a set of level
population equations with non-linear coefficients. Solution of this set of
coupled algebraic equations produces level populations that are self-consistent
with the line radiation they generate. Any desired level of accuracy can be
achieved by increasing the number of zones.

We introduce our new method in \S2. In \S3 we study the 2-level model in both a
semi-infinite atmosphere and finite slabs, presenting results and comparison
with SCP calculations. The new CEP method attains the exact solutions,
outperforming the SCP method by substantial margins. We present the equations
for multi-level systems in \S4 and include as an example an application to the
$^3$P system of \ion{O}{I}. Section 5 contains a discussion that, among other
things, covers various technical details.

\section{The New Technique}

Consider the transfer of a line with frequency $\nu_0$. The dimensionless line
profile is $\Phi(x)$, normalized so that $\int\Phi(x)dx = 1$, where $x = (\nu -
\nu_0)/\DnuD$ is the dimensionless frequency shift from line center and \DnuD\
is the Doppler width. We consider here only the case of $\Phi(x)$ that does not
change its shape throughout the source and frequency independent line source
function $S$. These assumptions are adopted to simplify the presentation. They
do not reflect inherent limitations of our new method.

\subsection                   {Radiative Transfer}

\begin{figure}
 \Figure{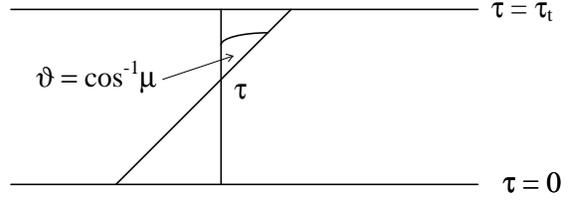}{0.9\hsize}
 \vskip0.2in
 \Figure{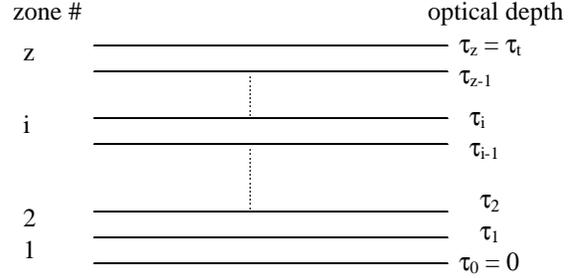}{0.9\hsize}
\caption{{\em Top}: Sketch of the slab geometry for the radiative transfer
problem.  {\em Bottom}: The partition of the slab into zones (see
\S\ref{sec:solution}).}
 \label{fig:geometry}
\end{figure}

For the geometry we adopt a plane-parallel slab whose physical properties vary
only perpendicular to the surface.  Optical depth at frequency $\nu$ along a
path orthogonal to the surface is $\tau_\nu = \tau\Phi(x)$, and $\tau$ can be
used as a coordinate that uniquely specifies locations in the slab (see figure
\ref{fig:geometry}). The optical depth along a ray slanted at $\theta =
\cos^{-1}\!\!\mu$ from normal is $\tau_\nu(\mu) = \tau\Phi(x)/\mu$, and the
intensity along the ray obeys the radiative transfer equation
\eq{\label{eq:rad_tran}
     \mu{dI_\nu(\tau, \mu)\over d\tau} = \Phi(x)[S(\tau) - I_\nu(\tau, \mu)]
}
The equation for the flux $F_\nu = 2\pi\int I_\nu\mu d\mu$ is obtained from
integration over angles. The overall line flux $F = \int F_\nu d\nu$ obeys at
every position in the slab
\eq{\label{eq:flux}
     {dF(\tau)\over d\tau} = 4\pi\DnuD[S(\tau) - \Jbar(\tau)]
}
where
\eq{\label{eq:Jbar}
    \Jbar(\tau) = \int {d\Omega\over4\pi} \int I_\nu(\tau,\mu) \Phi(x) dx
}
is the intensity averaged over both angles and line profile. Denoting by $\tot$
the overall optical thickness and accounting for the emission from both faces
of the slab, the line contribution to the cooling rate per unit area is
\eq{
    \Lambda = F(\tot) - F(0) \equiv 4\pi\DnuD\j
}
The line cooling factor \j\ is introduced for convenience when \DnuD\ is
constant in the slab. Integrating equation \ref{eq:flux} over $\tau$ yields
\eq{\label{eq:j}
               \j = \int_0^{\tot} S(\tau)p(\tau)d\tau
}
Here we introduced
\eq{\label{eq:p1}
   p(\tau) = 1 - {\Jbar(\tau)\over S(\tau)},
}
a quantity that has been called the net radiative bracket (Athay \& Skumanich
1971). From the formal solution of the radiative transfer equation,
\eq{\label{eq:p2}
   p(\tau) = 1 - {1\over 2S(\tau)}\int_0^{\tot}\!\!\! S(t)dt
                 \int_{-\infty}^{\infty}\!\!\!\Phi^2dx
                \int_0^1\!\!\! e^{-|\tau - t|\Phi/\mu}{d\mu\over\mu}
}
when there is no external radiation entering the slab.

\subsection                   {Level Populations}

Denote by $n_k(\tau)$, with $k = 1, 2$, the populations per sub-state of a
given transition at position $\tau$; that is, $n_k = N_k/g_k$ where $g_k$ is
the level degeneracy and $N_k$ is the overall level population. Then the line
source function is
\eq{\label{eq:S}
   S = {A_{21}\over B_{21}}{n_2\over n_1 - n_2}
}
where $A$ and $B$ are the Einstein coefficients of the transition. The
populations are obtained from steady-state rate equations of the form $\sum
R_{ij} = 0$. The term corresponding to exchanges between the transition levels,
separated by $E_{21} = h\nu_0$, is
\eq{\label{eq:rate1}
     R_{21} = -A_{21}n_2 - B_{21}\Jbar(n_2 - n_1)
              - C_{21}\!\!\left(n_2 - n_1 e^{-E_{21}\!/kT}\right)
}
where $C$ is the collision rate; exchanges with other levels have similar form
and are listed in \S\ref{sec:multi}.

\subsection                      {Solution}
\label{sec:solution}

The common approach of exact solution methods is to handle radiative transfer
and the level population distribution as two distinct problems, coupled through
the results each of them gives. The problem is initialized with populations
(and the corresponding source functions) obtained in some limiting case, e.g.,
thermal equilibrium. 
With these populations, radiative transfer (eq.\ \ref{eq:rad_tran}) is solved
for the intensity to determine \Jbar\ (eq.\ \ref{eq:Jbar}), which is then
plugged into the rate terms (eq. \ref{eq:rate1}) to determine new populations,
and so on. However, from eqs.\ \ref{eq:p1} and \ref{eq:S}, the rate term can be
written as
\eq{\label{eq:R21}
     R_{21} = -A_{21}n_2p - C_{21}\left(n_2 - n_1 e^{-E_{21}\!/kT}\right)
}
showing that the only radiative quantity actually needed for the calculation of
level populations at every position is the net radiative bracket $p(\tau)$;
given this factor we could compute the level populations that are consistent
with the radiation they produce without solving for the intensity. And as is
evident from equations \ref{eq:p2} and \ref{eq:S}, the factor $p(\tau)$ itself
can be computed from the level populations, again without solving  for the
intensity. Therefore, inserting $p(\tau)$ from equation \ref{eq:p2} into the
rate terms (eq.\ \ref{eq:R21}) produces {\em level population equations that
properly account for all the effects of radiative transfer without actually
calculating the intensity itself}; the radiative transfer equation has been
incorporated through its formal solution in equation \ref{eq:p2}.

A numerical solution of the resulting level population equations requires a
spatial grid, partitioning the source into zones such that all properties can
be considered uniform within each zone. The degree of actual deviations from
uniformity, and the accuracy of the solution, can be controlled by decreasing
each zone size through finer divisions with an increasing number of zones.
Figure \ref{fig:geometry} shows the slab partitioning into $z$ zones. The
$i$-th zone, $i = 1 \dots z$, occupies the range $\tau_{i-1} < \tau \le
\tau_{i}$, with $\tau_0=0$ and $\tau_z=\tau_t$. The optical depth between any
pair of zone boundaries is
\eq{\label{eq:tij}
    \t(i,j) = |\tau_i - \tau_j|
}
so that the optical thickness of the $i$-th zone is $\t(i,i - 1)$. The
temperature and collision rates are constant in the zone, and the corresponding
rate term for its (constant) level populations is
\eq{\label{eq:rate}
     R_{21}^i = -A_{21}n_2^ip^i
                - C_{21}^i\left(n_2^i - n_1^i e^{-E_{21}\!/kT_i}\right),
}
where the superscript $i$ is used as a zone label. The factor $p(\tau)$ varies
in the zone and has been replaced by a constant $p^i$ that should adequately
represent its value there, for example $p^i = \half\left[p(\tau_{i}) +
p(\tau_{i-1})\right]$ or $p^i = p\left(\half[\tau_{i} +\tau_{i-1}])\right)$.
There are no set rules for this replacement other than it must obey $p^i \to
p(\tau_{i})$ when $\t(i,i-1) \to 0$. We choose for $p^i$ the zone average
\eq{\label{eq:avg}
      p^i = {1\over\t(i,i-1)}\int_{\tau_{i-1}}^{\tau_i} p(\tau)d\tau
}
and this choice proved to be very successful in our numerical calculations.
From eq.\ \ref{eq:p2}, calculation of $p^i$ requires an integration over the
entire slab, which can be broken into a sum of integrals over the zones. In
each term of the sum, the zone source function can be pulled out of the
$\tau$-integration so that
\eqarray{
    p^i &= &1 - {1\over2\t(i,i-1)S^i}
         \sum_{j = 1}^z S^j \times \non
         &&\int_{\tau_{i - 1}}^{\tau_{i}}\!\!\!d\tau
         \int_{\tau_{j - 1}}^{\tau_{j}}\!\!\! dt
                 \int_{-\infty}^{\infty}\!\!\!\Phi^2dx
                \int_0^1\!\!\! e^{-|\tau - t|\Phi/\mu}{d\mu\over\mu}
}
The remaining integrals can be expressed in terms of common functions. Consider
for example
\eqarray{
    \beta^i &=& 1 - {1\over2\t(i,i-1)}\times \non
         &&\int_{\tau_{i - 1}}^{\tau_{i}}\!\!\!d\tau
          \int_{\tau_{i - 1}}^{\tau_{i}}\!\!\!dt
                 \int_{-\infty}^{\infty}\!\!\!\Phi^2dx
                \int_0^1\!\!\! e^{-|\tau - t|\Phi/\mu}{d\mu\over\mu},
}
the contribution of zone $i$ itself to $p^i$. It is straightforward to show
that $\beta^i = \beta(\t(i,i - 1))$, where
\eq{\label{eq:beta}
  \beta(\tau) = {1\over\tau}\int_0^\tau \!\!\!dt
                            \int_{-\infty}^{\infty}\!\!\!\Phi(x)dx
                            \int_0^1 \!\!\!d\mu\, e^{-t\Phi(x)/\mu}
}
This function was first introduced by Capriotti (1965); it is the probability
for photon escape from a slab of thickness $\tau$, averaged over the photon
direction, frequency and position in the slab. The contribution of zone $j \ne
i$ to the remaining sum can be handled similarly, and the final expression for
the coefficient $p^i$ is
\eq{\label{eq:pi}
     p^i = \beta^i + {1\over\t(i,i-1)}\!\!
     \sum_{\stackrel{j = 1}{j \ne i}}^z \!\! {S^j\over S^i} M^{ij}
}
where
\eq{\label{eq:M}
    M^{ij} = -\frac12(\a(i,j) - \a(i-1,j) - \a(i,j-1) + \a(i-1,j-1))
}
and where $\a(i,j) = \t(i,j)\beta(\t(i,j))$. The quantity $\a(i,j)$ obeys
$\a(i,j) = \a(j,i)$ and $\a(i,i) = 0$, therefore $M^{ij} = M^{ji}$ and $M^{ii}
= \a(i,i-1)$.\footnote{Since $\beta^i = M^{ii}/\t(i,i-1)$, the first term could
be incorporated into the sum in eq.\ \ref{eq:pi} as the $j = i$ term.} The
first term in the expression for $p^i$ is the average probability for photon
escape from zone $i$, reproducing one of the common variants of the escape
probability method in which the whole slab is treated as a single zone (e.g.,
Krolik \& McKee 1978). The subsequent sum describes the effect on the level
populations in zone $i$ of radiation produced in all other zones. Each term in
the sum has a simple interpretation in terms of the probability that photons
generated elsewhere in the slab traverse every other zone and get absorbed in
zone $i$, where their effect on the level populations is similar to that of
radiation external to the slab (see appendix \ref{sec:external}).

Inserting the coefficients $p^i$ from eq.\ \ref{eq:pi} into the rate terms
(eq.\ \ref{eq:rate}) in every zone produces a set of non-linear algebraic
equations for the unknown level populations $n_k^i$. The procedure was outlined
here only for the diffuse radiation of a single transition; we describe the
extension to multi-levels in \S\ref{sec:multi} and the inclusion of external
radiation in appendix \ref{sec:external}. {\em Solution of these equations
yields the full solution of the line transfer problem by considering only level
populations};\footnote{Apruzese et al (1980) proposed somewhat similar
equations. They based their arguments on probabilistic reasoning and did not
offer a formal derivation. We thank P.\ Lockett for bringing this to our
attention.} the computed populations are self-consistent with their internally
generated radiation even though the radiative transfer equation is not handled
at all. Once the populations are found, radiative quantities can be calculated
in a straightforward manner from summations over the zones. The emerging
intensity at direction $\mu$ is
\eq{
    I_\nu(\tot,\mu) = \sum_{i = 1}^z
     \left(e^{-\tau^{z,i}\Phi/\mu} -  e^{-\tau^{z,i - 1}\Phi/\mu}\right)S^i.
}
The flux density emerging from each face of the slab obeys
\eqarray{
    F_\nu(\tot) &=& 2\pi\sum_{i = 1}^z
    \left[E_3(\tau^{z,i}\Phi) - E_3(\tau^{z,i - 1}\Phi)\right] S^i \non
    -F_\nu(0)  &=& 2\pi \sum_{i = 1}^z
    \left[E_3(\t(i - 1,0)\Phi) - E_3(\t(i,0)\Phi)\right] S^i
}
where $E_3$ is the third exponential integral (e.g., Abramowitz \& Stegun
1972). The line cooling coefficient is
\eq{
    \j = \half\sum_{i = 1}^z
 \left(\alpha^{i,0} - \alpha^{i-1,0} - \alpha^{z,i} + \alpha^{z,i-1}\right)S^i.
}

The solution method just described is exact --- the discretized equations are
mathematically identical to the original ones when $\t(i,i-1) \to 0$ for every
$i$. As is usually the case, the only approximation in actual numerical
calculations is the finite size of the discretization, i.e., the finite number
of zones. A desired accuracy is achieved when\textbf{,} upon further division\textbf{,} the
relative change in all level populations is smaller than the prescribed
tolerance.


\begin{figure}
 \centering\leavevmode
 \includegraphics[width=\hsize,clip]{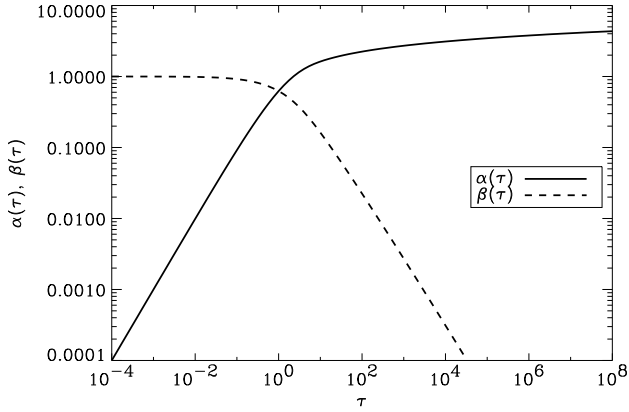}
\caption{Plots of the functions $\beta$ (see eq.\ \ref{eq:beta}) and $\alpha =
\tau\beta$ (eq.\ \ref{eq:alpha}).}
 \label{fig:alpha_beta}
\end{figure}


\subsection{Numerical Implementation}

The level populations of all zones are described by a set of non-linear
algebraic equations. The equations are readily solved by the Newton method,
which utilizes the Jacobian of the set. Since the dependence on the unknown
variables is explicit in all the rate terms, the Jacobian can be computed from
analytic expressions. The functions $\beta$ (see eq.\ \ref{eq:beta}), $\alpha =
\tau\beta$ and their derivatives are conveniently calculated from the
representations
 \newpage
\eqarray{\label{eq:alpha}
 \alpha(\tau)  &=& \int_{-\infty}^{\infty}dx \{\half - E_3[\tau\Phi(x)]\} \non
 \alpha'(\tau) &=& \int_{-\infty}^{\infty} \Phi(x) E_2[\tau\Phi(x)]dx
}
where $E_2$ is the second exponential integral. Figure \ref{fig:alpha_beta}
plots $\beta$ and $\alpha$. While $\beta$ is a monotonically decreasing
function, $\alpha$ is monotonically increasing and its asymptotic behavior when
$\tau \to \infty$ is $\alpha \sim \sqrt{\ln\tau}$. This divergent behavior does
not pose any problems because $\alpha$ cancels to first order in the thickness
of the zones in the linear combinations defining $M^{ij}$  (eq.\ \ref{eq:M}).
Only the second order terms, involving the second derivative $\alpha''$,
survive.

Since our aim is to explore the intrinsic accuracy of our new method, the
integrals in eq.\ \ref{eq:alpha} were computed repeatedly with an 80 points
Gaussian quadrature to ensure that these integrations do not compromise the
precision of the outcome. The $E_n$ functions were evaluated with a rapidly
convergent series from Press et al. (1986). The integration range was truncated
at $x = \pm7$, which we have verified is sufficient in all cases thanks to the
rapid decrease with $x$ of the integrands.

In order to test the new method, the radiative transfer problem was also solved
using the ALI method for comparison. The technique is based on a modified
$\Lambda$ iteration in which the statistical equilibrium equations are
linearized via the Rybicky \& Hummer (1992) preconditioning scheme. The method
also takes advantage of an operator splitting scheme by introducing an
approximate operator $\Lambda^*$, the diagonal of the exact operator $\Lambda$
in the formal solution $\Jbar = \Lambda[S]$ of the radiative transfer equation.
It has been shown that the introduction of this operator leads to an optimal
balance between the convergence rate and computing time per iteration (Olson,
Auer \& Buchler 1986; Carlsson 1991). The ALI calculations presented in this
paper utilize a formal solver based on the short-characteristics scheme with
parabolic precision (Olson, Auer \& Buchler 1986), currently considered the
method of choice for complicated line transfer problems (e.g., Kunasz \& Auer
1988; Auer, Fabiani Bendicho \& Trujillo Bueno 1994; van Noort, Hubeny \& Lanz
2002; Fabiani Bendicho 2003). With this SCP method, equation \ref{eq:rad_tran}
was solved for many frequencies and ray inclinations, and the mean intensity
computed from angular and frequency integrations (eq.\ \ref{eq:Jbar}) by
numerical quadratures.\footnote{It is interesting to note that the calculation
of the mean intensity has also been done using Monte Carlo techniques (see,
e.g., van Zadelhoff et al 2002 and references therein).} To ensure the high
precision required in this comparative study, the angular integration was done
with a Gaussian quadrature with 24 points in the variable $\mu$. The frequency
integrals were done with trapezoidal integration extending to $x = \pm 4$ with
33 frequency points, which we have verified yields the desired precision.

We proceed now to present solutions and comparisons of the newly developed CEP
method with the SCP method for a number of standard problems.
In all the examples we employ uniform physical conditions and the Doppler shape
for the line profile, $\Phi = \pi^{-1/2}e^{-x^2}$; note that the line center
optical depth is then $\tau_0 = \tau/\sqrt{\pi}$.

\section{2-level Atom}

In the two-level problem, the steady-state rate equation $R_{21}$ = 0 (eq.
\ref{eq:rate1}) yields the familiar expression for the source function
\eq{\label{eq:S2a}
    S = (1 - \epsilon)\Jbar + \epsilon B(T)
}
where $B$ is the Planck function and where
\eq{\label{eq:epsilon}
    {\epsilon\over 1 - \epsilon}
      = {C_{21}\over A_{21}}\left(1 - e^{-E_{21}\!/kT}\right)
      \equiv {N\over\Ncr}
}
Here $N$ is the density of the collision partners and \Ncr\ the standard
critical density with a slight modification that incorporates the Boltzmann
factor correction. Replacing \Jbar\ with $p$ (eq.\ \ref{eq:p1}), the equation
for the source function becomes
\eq{\label{eq:S2}
    (1 + \eta p)S = B,        \qquad \qquad \hbox{where}\qquad
    \eta = {\Ncr\over N};
}
this result also follows directly from eq.\ \ref{eq:R21} with $R_{21}$ = 0.
When the 2-level problem is formulated with optical depth as the independent
variable, it is fully characterized by the two input quantities $B(T)$ and
$\epsilon$ (or, equivalently, $\eta$) specified as functions of $\tau$. There
is no need to specify intrinsic properties of the transition such as, for
example, $E_{21}$ or $A_{21}$. Instead of solving for the population of each of
the two levels, this single equation for the unknown $S$ provides the complete
solution of the problem.

Dividing the slab into zones, the rate equation $R_{21}^i$ = 0 (eq.\
\ref{eq:rate}) produces a similar expression for the $i$-th zone,
\eq{
    S^i + \eta^ip^iS^i = B(T^i),
}
with $\eta^i$ and $T^i$ corresponding to the physical conditions in the zone.
Inserting the expression for $p^i$ from equation \ref{eq:pi}, the CEP set of
equations for the unknown $S^i$ is
\eq{\label{eq:S2CEP}
      \left(1 + \eta^i\beta^i\right)S^i
    + {\eta^i\over\t(i,i - 1)}\sum_{\stackrel{j = 1}{j \ne i}}^z M^{ij}S^j
    = B(T^i)
}
Since the factors $\beta^i$ and $M^{ij}$ depend only on optical depth, they are
independent of the unknown variables (the zone source functions $S^i$)  in this
case. Therefore, {\em the CEP technique transforms the two-level problem to a
set of linear equations}. This is a reflection of the linear relation between
$I_\nu$ and $S$ in eq.\ \ref{eq:rad_tran} that is maintained when the complete
problem is handled in terms of optical depth as the independent variable. The
CEP formulation produces directly the explicit linear equations in this case.

We proceed now with solutions for semi-infinite atmospheres and
finite-thickness slabs with constant physical conditions. When the temperature
is constant, $B(T)$ merely sets the intensity scale and only the dependence on
$\epsilon$ need be studied.

\subsection{Semi-infinite Atmosphere}
\label{sec:atmosphere}


\begin{figure}
 \centering\leavevmode
\def\dir{source_function_semiinf}
 \includegraphics[width=\hsize,clip]{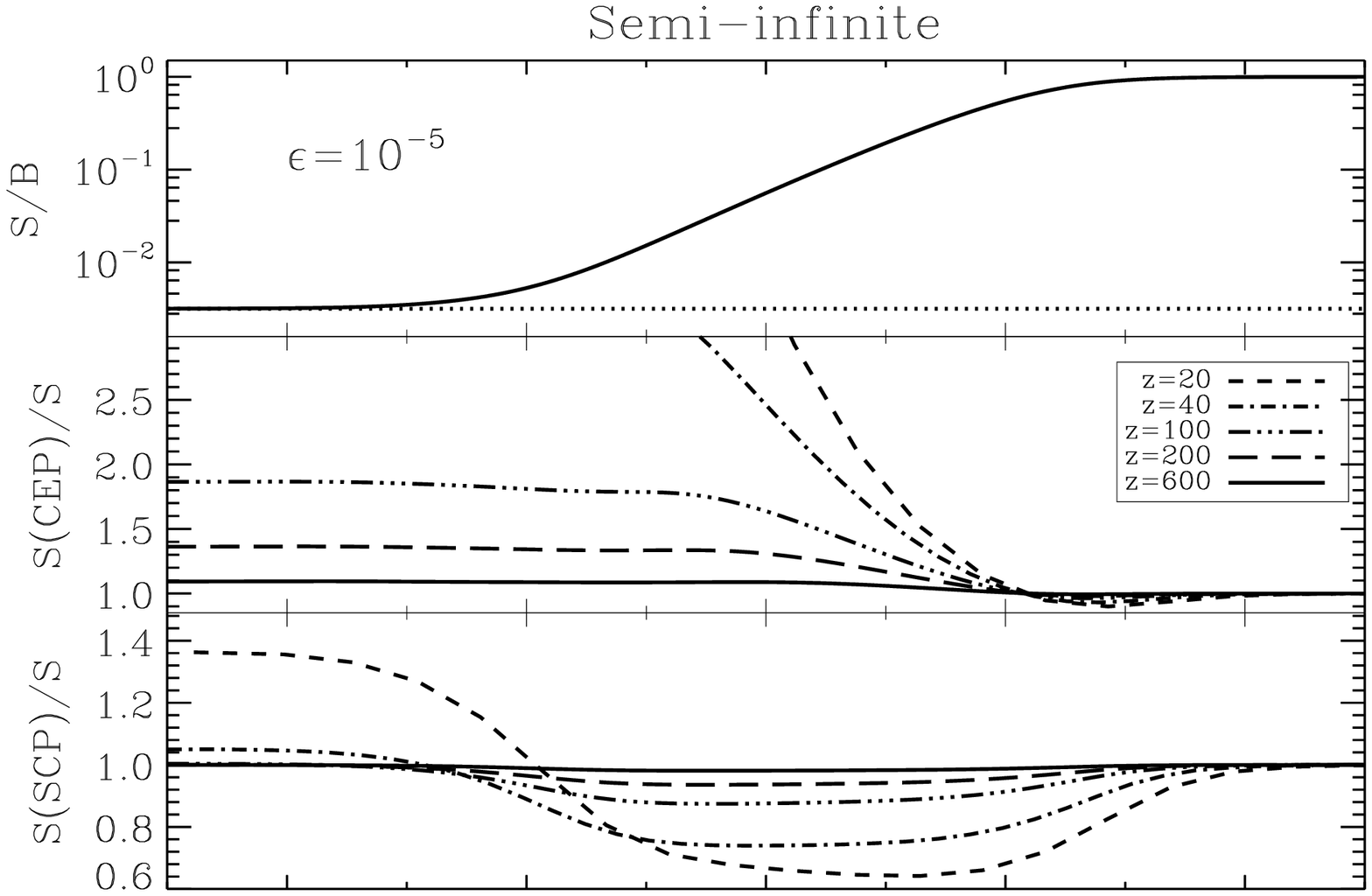}
 \includegraphics[width=\hsize,clip]{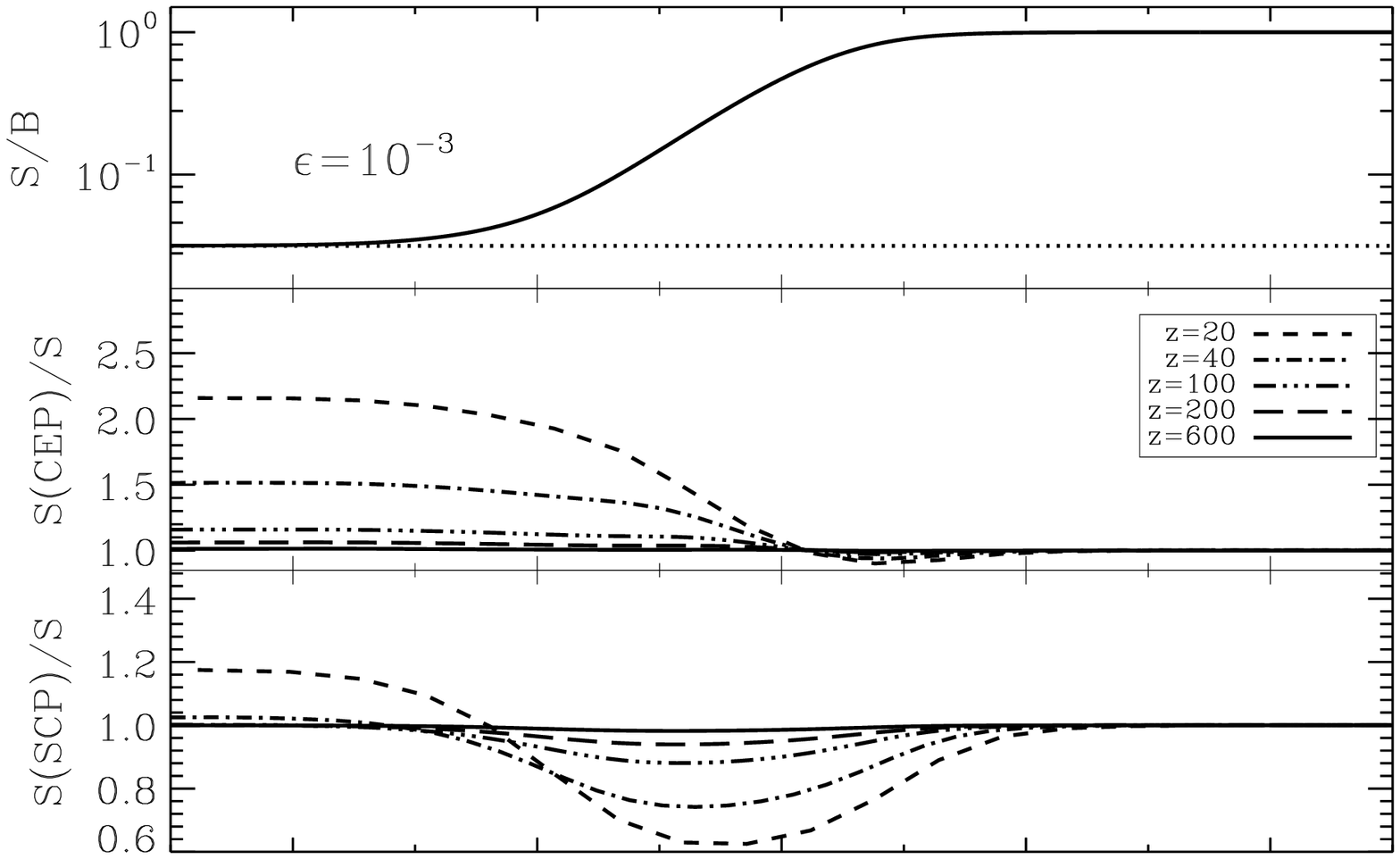}
 \includegraphics[width=\hsize,clip]{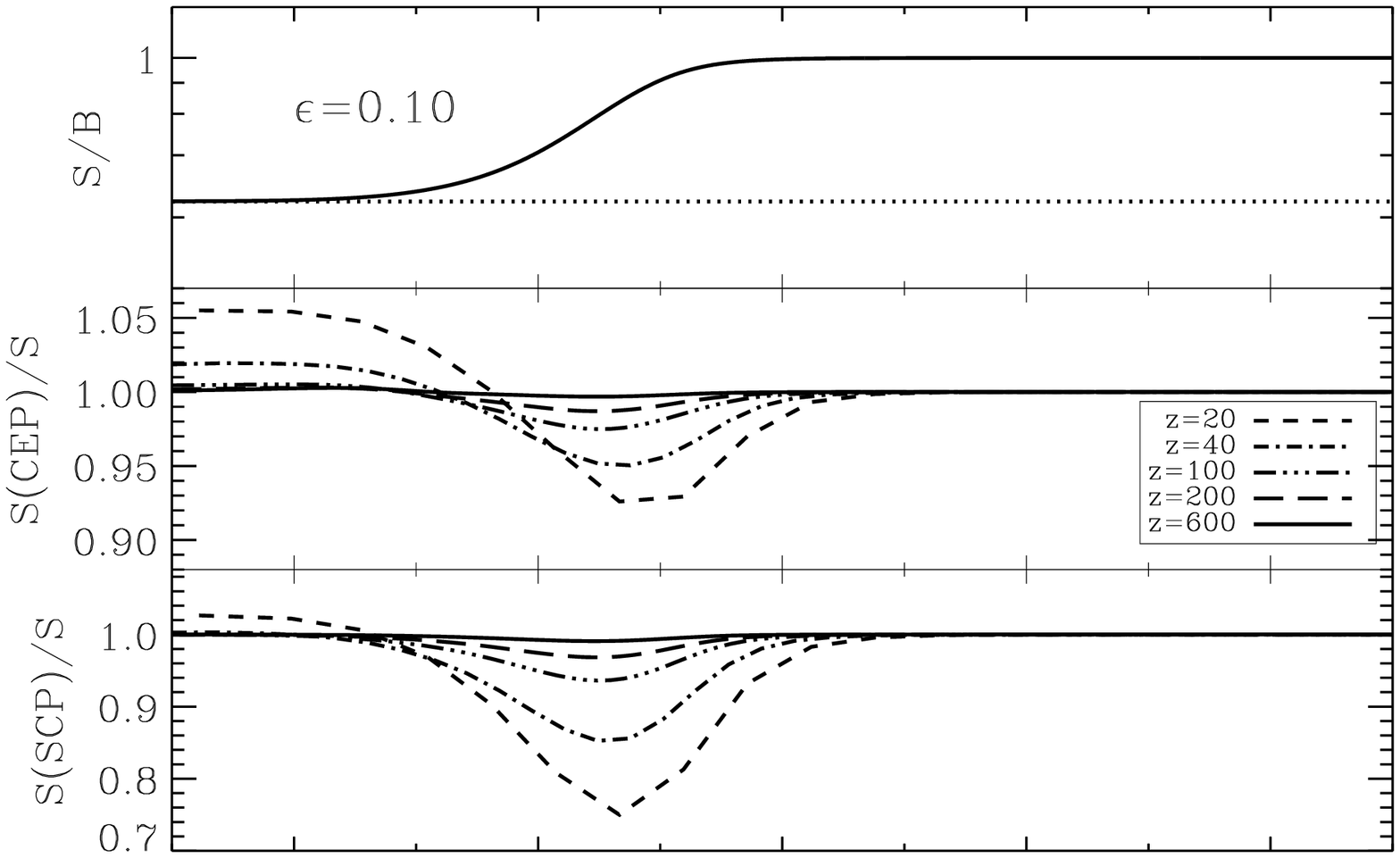}
 \includegraphics[width=\hsize,clip]{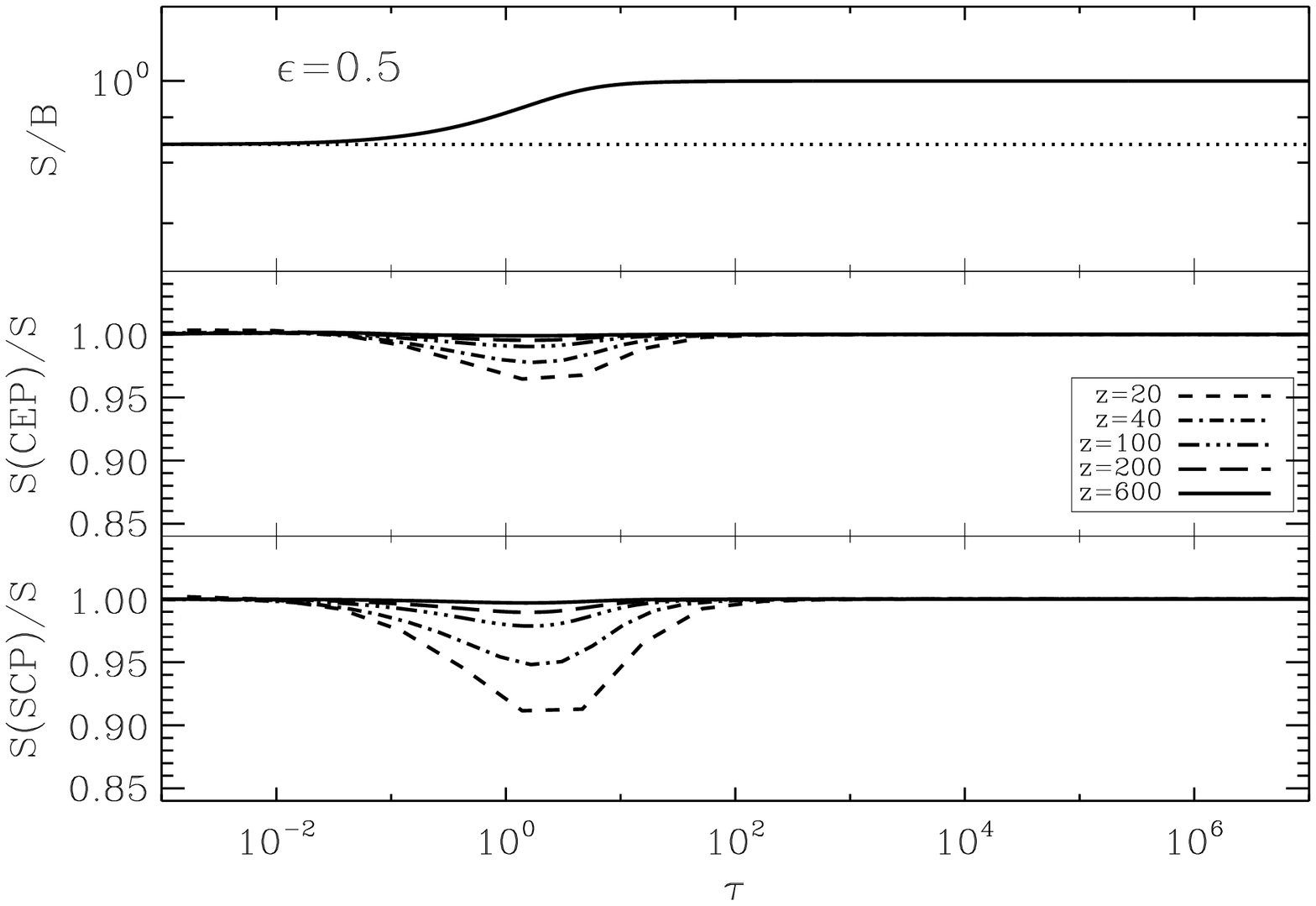}
\caption{The two-level model with various values of $\epsilon$ (eq.\
\ref{eq:epsilon}) in a semi-infinite atmosphere. The top panel of each plot
shows the variation of the source function with depth into the atmosphere. The
two other panels show the convergence to the exact solution as the number of
zones $z$ is increased for the CEP and SCP methods. Note the change in scale of
the vertical axis between different panels.}
 \label{fig:atmosphere}
\end{figure}



\begin{table}
\centering
\begin{tabular}{r rrr rrr}
  \hline
  \hline
   Zones &&\multicolumn{2}{c}{Time} && \multicolumn{2}{c}{\% error} \\
         &&  SCP  & CEP  &&  SCP & CEP   \\
   \hline
     20  &&  0.39 & ---  && 36.3 & 103.6 \\
     40  &&  1.10 & ---  && 23.9 &  45.7 \\
    100  &&  4.39 & .006 && 10.9 &  14.0 \\
    200  &&  11.6 & .089 &&  5.5 &   5.4 \\
    600  &&  44.9 & 1.70 &&  1.6 &   1.2 \\
\end{tabular}
\caption{Runtime (in seconds, on the same computer) required by the SCP and CEP
methods to solve an atmosphere with $\epsilon = 10^{-3}$ for the number of
zones listed in the first column; in the SCP method this corresponds to the
number of grid points. Omitted entries were too short for meaningful timing.
The listed error is the percent deviation from the result of an SCP calculation
with 3,000 zones.
}
\label{table:atmosphere}
\end{table}

We start with the classical problem of a stellar atmosphere, where $\tau$
measures distance from the surface and $\tot \to \infty$. The source function
is subject in this case to the exact limits
\eq{
    S \to B\times\cases{\sqrt{\epsilon} & when $\tau \to 0$         \cr
                                                                    \cr
                        1               & when $\tau \gg 1/\epsilon$  }
}
(e.g. Avrett \& Hummer 1965). In order to capture both limit behaviors we model
the atmosphere as a slab divided logarithmically into $z$ zones that cover ten
orders of magnitude in optical depth from $\tau = \E{-3}$ to $\tot = \E7$, with
the latter serving as a proxy for the atmospheric interior. The two faces of
the slab are a-priori identical. When the radiative transfer equation is part
of the calculation, this two-sided symmetry is broken by the boundary condition
$I_\nu(\tau = \tot,\mu) = 0$, which introduces a radiation sink at the
$\tot$-boundary. This is the case in ALI methods, including SCP. The CEP
method, on the other hand, does not involve the radiation at all and thus
cannot invoke boundary conditions to differentiate between the slab two faces.
Instead, this is accomplished by the logarithmic division that starts at one
end, and the great disparity that this introduces between photon escape from
the two sides. The semi-infinite atmosphere could also be mimicked by doubling
the slab with its mirror image and considering the source function only between
one surface and the mid-plane. We have verified that the results of
calculations with the two approaches are practically identical. In order to
compare the CEP method with SCP under identical conditions we present the
results for logarithmic divisions increasing toward the slab surface at $\tot$.

Figure \ref{fig:atmosphere} shows the results for some representative models,
ranging from $\epsilon$ = \E{-5} ($N = \E{-5}\Ncr$) to $\epsilon$ = 0.5 ($N =
\Ncr$). By example, the \ion{Ca}{II} H line can be modeled in a 5000 K
atmosphere with $\epsilon$ = 3.65\x\E{-5} (e.g., Avrett \& Loeser 1987 and
references therein). The top panel of each plot shows the solution obtained
with the SCP method with 3,000 zones, displaying the proper limit behavior at
both ends of the optical depth axis. The CEP method attains these solutions
with a sufficient number of zones, validating our new technique. But the
convergence with $z$ is quite different for the two methods.

At the surface, the SCP method is close to the exact solution already at $z$ =
20 (only two zones per logarithmic decade) in all cases; the deviation is less
than 40\% at $\epsilon = \E{-5}$ and decreases further as $\epsilon$ increases.
Increasing $z$ brings rapid convergence. In contrast, deep inside, the rate of
convergence is much more moderate. Furthermore, when $\epsilon$ increases, both
the magnitude of deviations and the rate of convergence around $\tau \sim 10$
remain almost the same for all $\epsilon \le 0.1$. The behavior at both ends
reflects the short characteristics nature of the method, in which only nearby
regions are coupled, and the fact that the radiative transfer equation is
always solved. Radiative transfer effects are minimal at small $\tau$, which is
why the method attains easily the exact solution near the surface. But the
effects are significant at the optical depths where the transition to
thermalization occurs, the radiative transfer equation must be repeatedly
solved and the convergence in these regions is hardly improved by the increase
in collision rates as long as $N$ remains sub critical.

In an almost mirror behavior, at the surface the CEP method deviates from the
exact solution by more than factor 2 at $z < 100$ when $\epsilon$ is small, and
its convergence rate to the exact solution is slow there. However, deep inside
the atmosphere, the deviations are actually smaller than at the surface.
Moreover, when $\epsilon$ increases, the deviations decrease everywhere. At
$\epsilon$ = 0.1, the CEP method is within 7\% of the exact solution everywhere
already at $z$ = 20; in contrast, this accuracy is attained by the SCP method
only at $z$ = 100. These properties are readily understood from the CEP
formalism. Since the level population equations couple the entire atmosphere,
the surface layers are affected by the behavior deep inside. And because the
radiative transfer equation is avoided altogether, the method takes full
advantage of the thermalization that approaches the surface when $\epsilon$
increases.

Performance statistics for the two methods are summarized in Table
\ref{table:atmosphere} for the case $\epsilon$ = \E{-3}; the statistics for
other cases show similar trends. The CEP technique attains the solution much
faster than the SCP method in all cases.


\begin{figure}
 \centering\leavevmode
 \includegraphics[width=\hsize,clip]{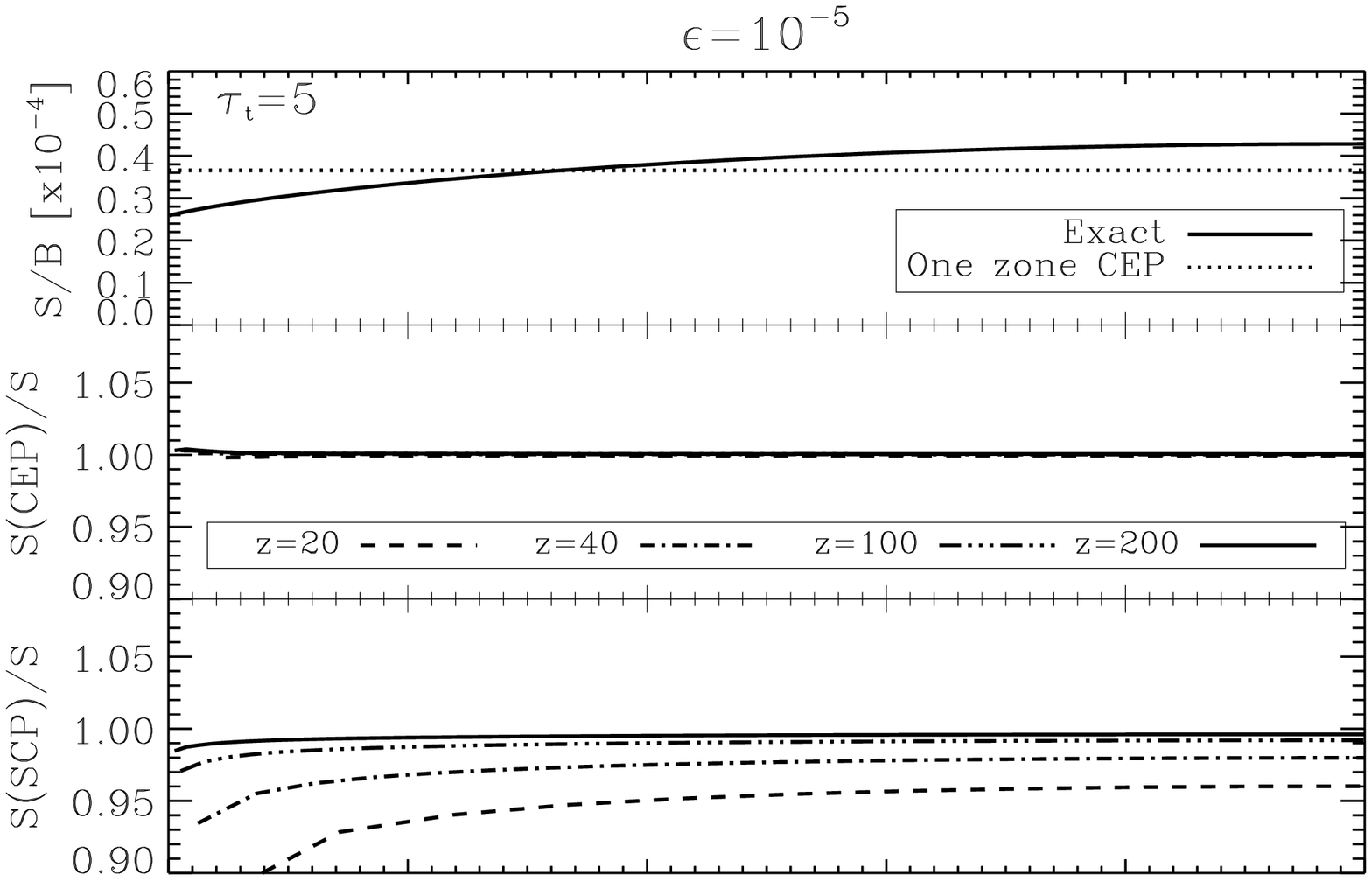}
 \includegraphics[width=\hsize,clip]{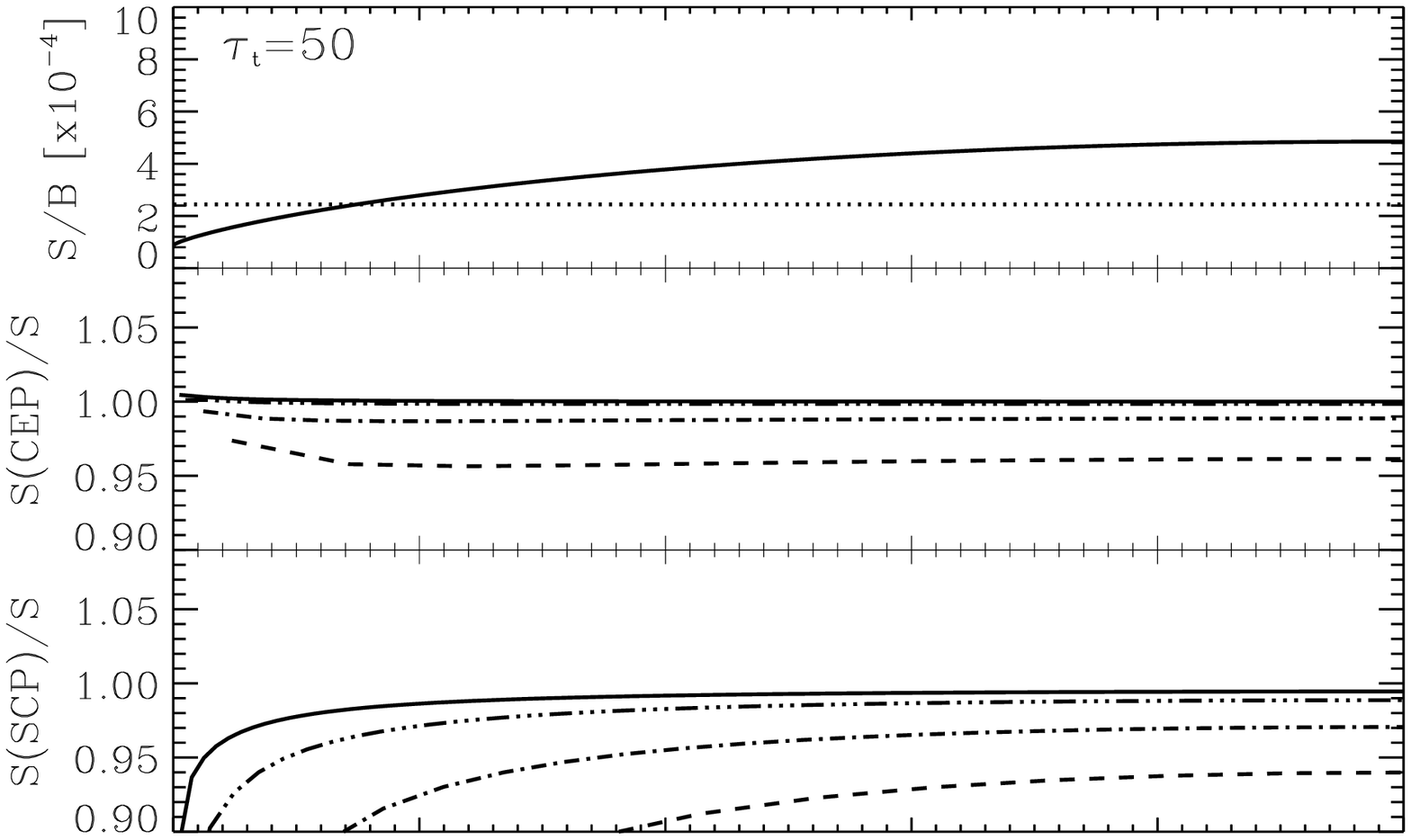}
 \includegraphics[width=\hsize,clip]{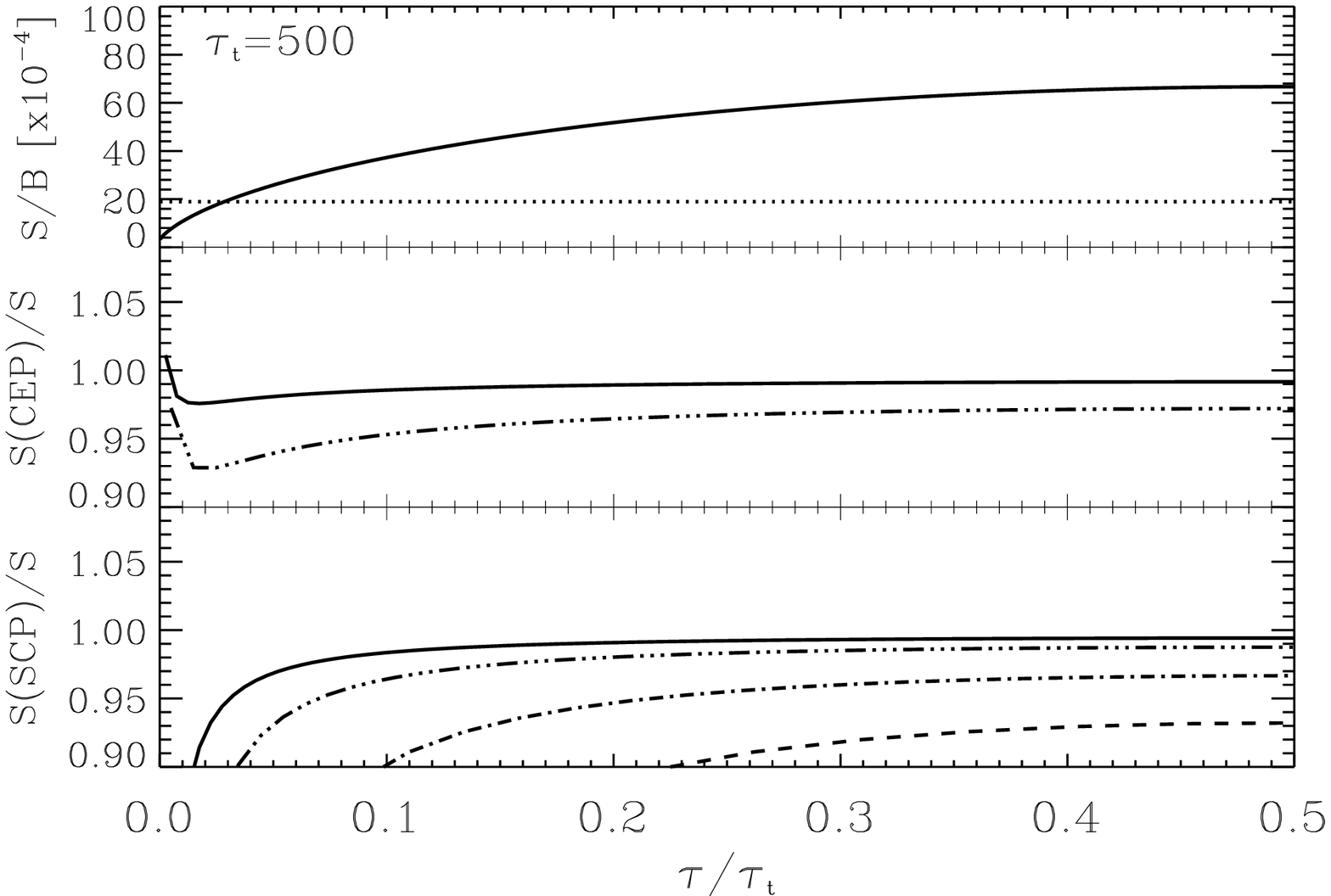}
\caption{The two-level model with $\epsilon$ = \E{-5} in slabs with overall
optical thickness $\tot$. The top panel of each plot shows the variation of the
source function from the surface to the slab mid-plane in the exact solution
and in a single-zone CEP calculation. The two other panels show the convergence
to the exact solution with the number of zones $z$ for the CEP and SCP
methods.}
 \label{fig:slab_S_epsE-5}
\end{figure}



\begin{figure}
 \centering\leavevmode
 \includegraphics[width=\hsize,clip]{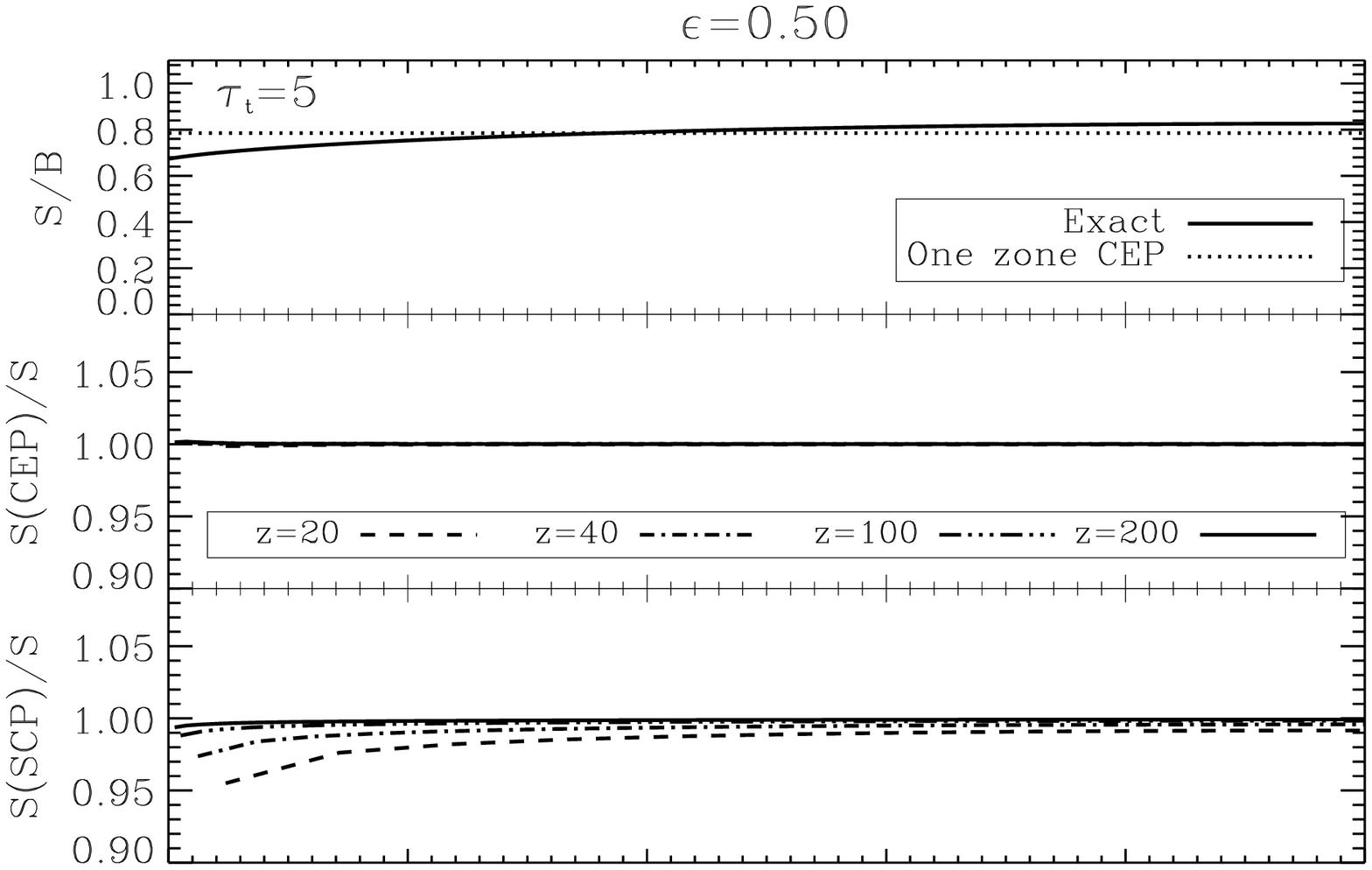}
 \includegraphics[width=\hsize,clip]{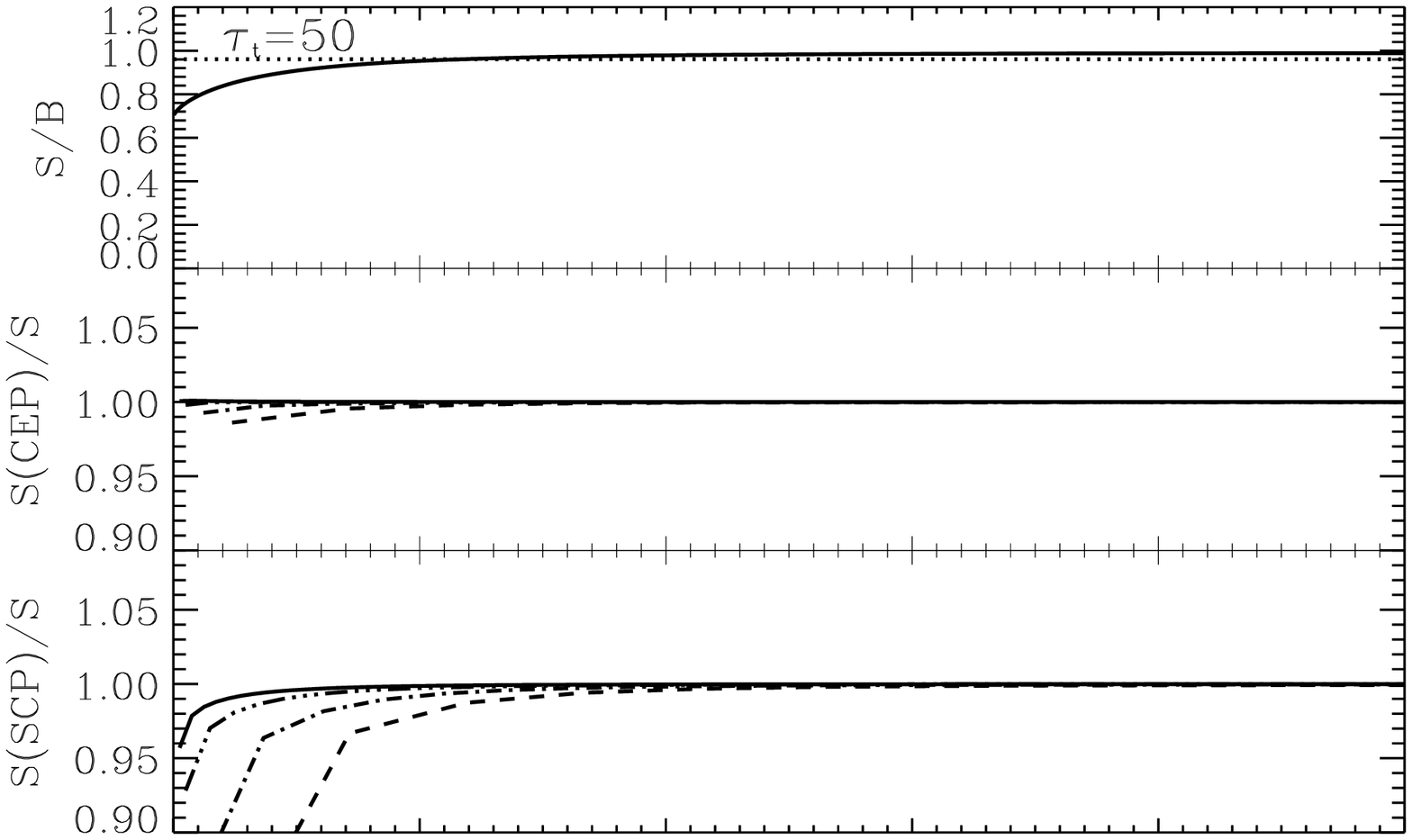}
 \includegraphics[width=\hsize,clip]{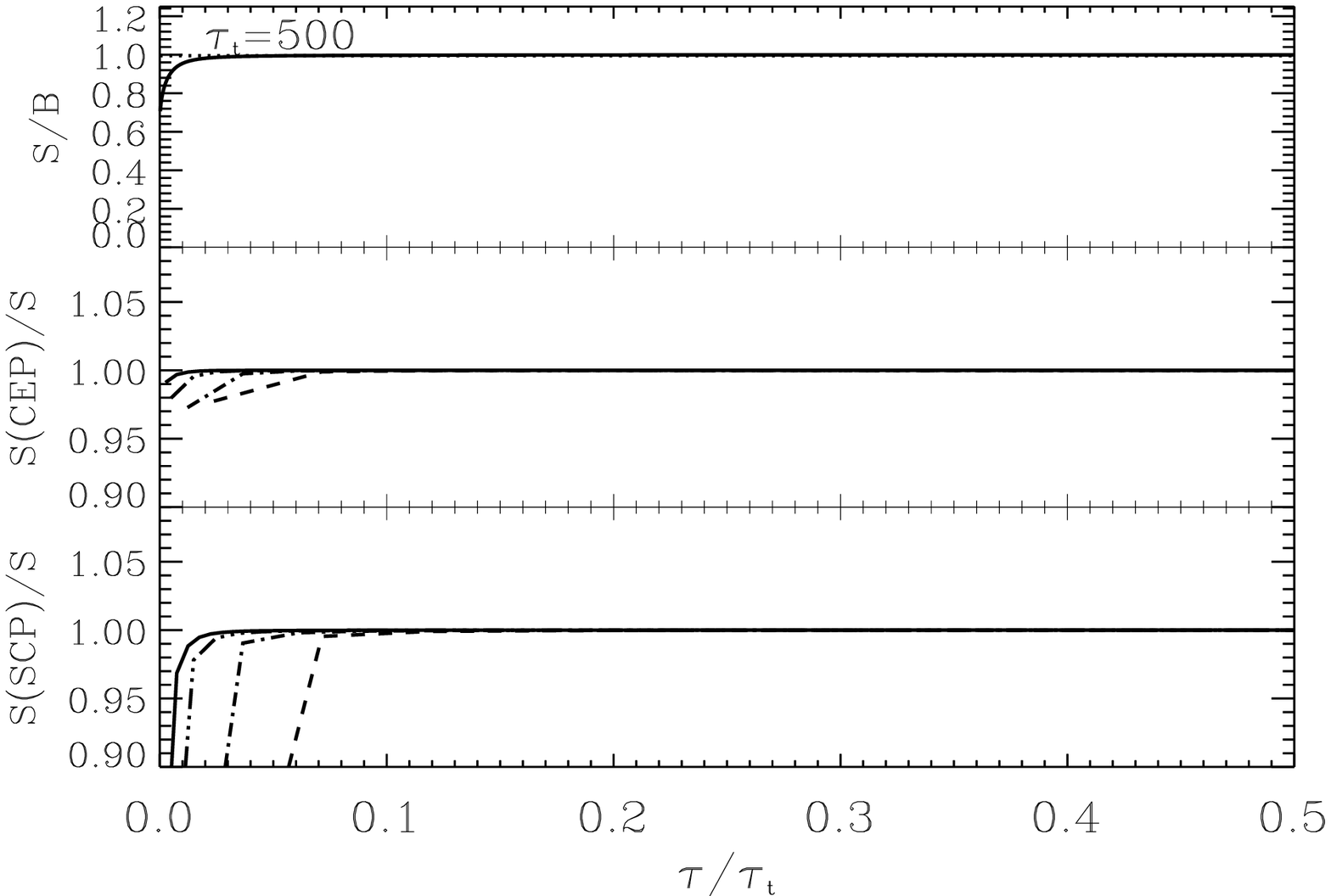}
\caption{Same as figure \ref{fig:slab_S_epsE-5}, only $\epsilon$ = 0.5}
 \label{fig:slab_S_eps0.5}
\end{figure}


\subsection{Finite-thickness Slabs}

One of the main coolants of photodissociation regions (PDR) is the 158 \mic\
line of \ion{C}{II}, whose emission is often modeled with a simple escape
probability approximation of the two-level system (e.g., Tielens and Hollenbach
1985). When this approach employs the Capriotti expression for the escape
probability (eq.\ \ref{eq:beta}), it is identical to a CEP calculation with
only one zone. For comparison with exact solutions, we include such single-zone
CEP calculations in the results presented here. The numerical calculations
employ $z$ zones of equal thickness.

Figure \ref{fig:slab_S_epsE-5} shows the variation of the source function
inside slabs of various optical thickness for $\epsilon = \E{-5}$. The
displayed behavior is representative of all $N \ll \Ncr$ cases. The results of
single-zone CEP calculations provide reasonable approximations at small $\tot$,
but become poorer as the variation range of the source function gets wider with
increasing $\tot$. However, with only 20 zones the CEP results are within 1\%
of the exact solution everywhere when $\tot \le 10$, 4\% when $\tot \le 50$ and
10\% when $\tot \le 100$. An accuracy better than 10\% is always achieved when
the optical thickness of each zone is $\la$ 5. In contrast, the SCP method does
not reach this level of accuracy near the surface of a $\tot$ = 500 slab even
with 200 zones; as a grid- rather than zone-based method, it attempts to
resolve the surface structure even when that is not required. Furthermore, SCP
calculations require a large number of divisions even at moderate optical
thickness; when $\tot$ = 10, a 10\% accuracy requires 100 zones. The reason, as
noted above, is that the equation of radiative transfer must be solved
repeatedly; the approach to thermal equilibrium of level populations deep
inside the slab does not alleviate this need, and large optical depths dictate
a large number of zones.

Since the CEP technique employs only level populations it takes full advantage
of level thermalization. The difference with standard methods in the case of
$\epsilon$ = 0.5 ($N$ = \Ncr), shown in figure \ref{fig:slab_S_eps0.5}, is
striking. Already with one zone, CEP calculations produce acceptable results
inside every slab, even with $\tot$ as large as 500; 20 zones suffice for 3\%
accuracy everywhere. In contrast, to achieve 10\% accuracy, SCP requires 100
zones at a moderate $\tot$ = 50, and even 200 zones are insufficient when
$\tot$ = 500. The zone thicknesses in this case is $\tau$ = 2.5, enough to
challenge numerical solutions of the radiative transfer equation that SCP must
perform.

Table \ref{table:slab} summarizes the performance statistics in one
representative case. The CEP method outperforms SCP by an even larger margin
than in the case of an atmosphere.


\begin{table}
\centering
\begin{tabular}{r| rr rr rr}
\hline \hline
    & \multicolumn{2}{c}{Time}
    & \multicolumn{2}{c}{\% error -- S}
    & \multicolumn{2}{c}{\% error -- \j}
    \\
    \hline
   Zones &  SCP  & CEP    &  SCP  &  CEP  &  SCP  &  CEP  \\
    \hline
       1 &  0.03 &  ---   &  99.5 &  55.5 &  98.6 &  25.0 \\
      10 &  0.60 &  ---   &  93.3 &  31.0 &  28.2 &  13.2 \\
      20 &  1.02 &  ---   &  88.2 &  23.0 &  17.5 &  7.94 \\
      40 &  3.29 &  ---   &  79.8 &  15.5 &  10.1 &  4.01 \\
     100 &  12.6 &  .033  &  62.5 &  6.47 &  3.94 &  1.21 \\
     200 &  28.1 &  .085  &  46.4 &  2.22 &  1.60 &  0.40 \\
\end{tabular}
\caption{Performance comparison of the SCP and CEP methods, similar to Table
\ref{table:atmosphere}, for a slab with $\epsilon = 10^{-3}$  and $\tot$ = 500.
The listed errors include the percent deviations of both the source function
and the line cooling coefficient from the exact results.
}
\label{table:slab}
\end{table}


\begin{figure}
 \centering\leavevmode
 \includegraphics[width=\hsize,clip]{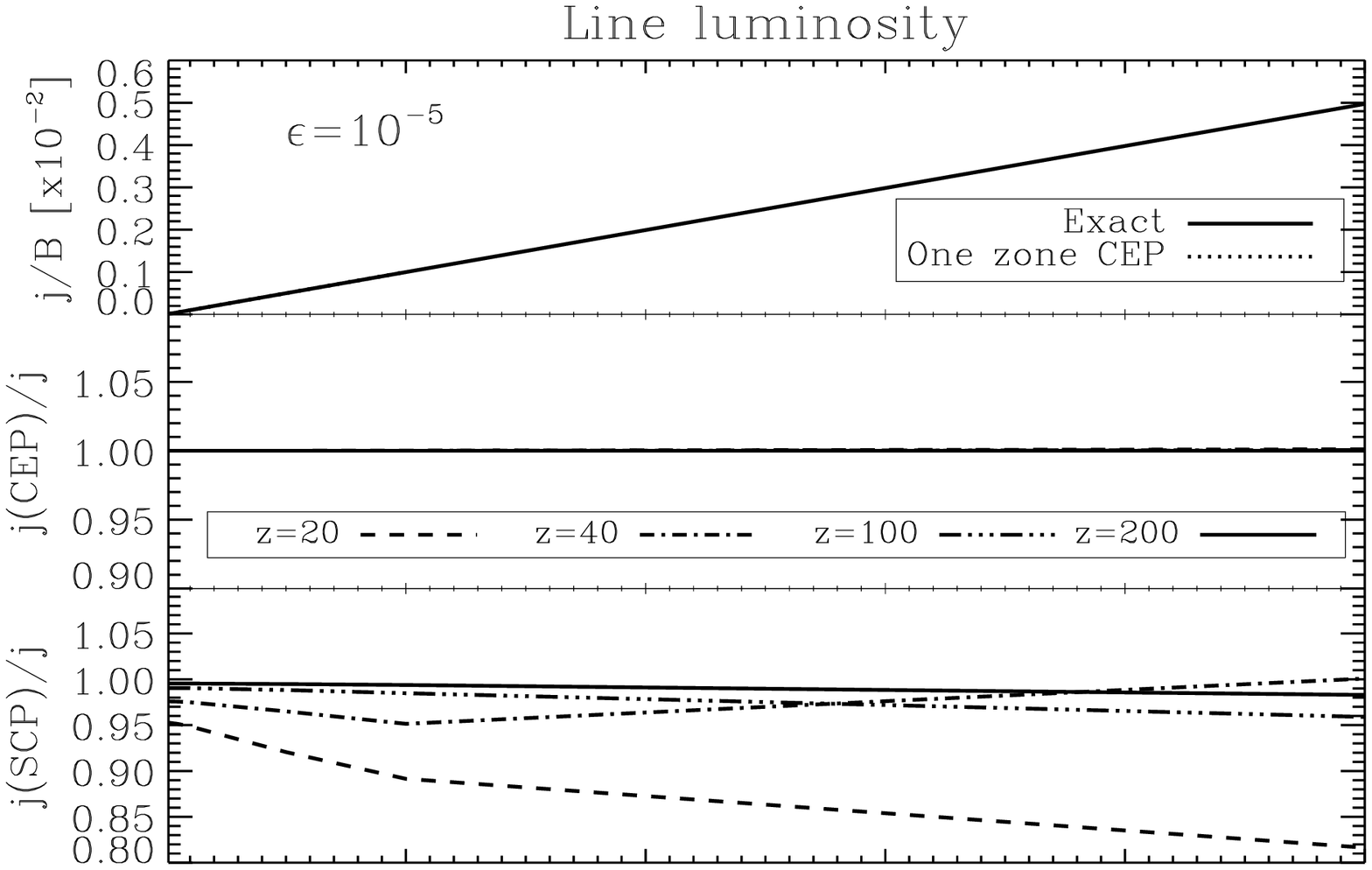}
 \includegraphics[width=\hsize,clip]{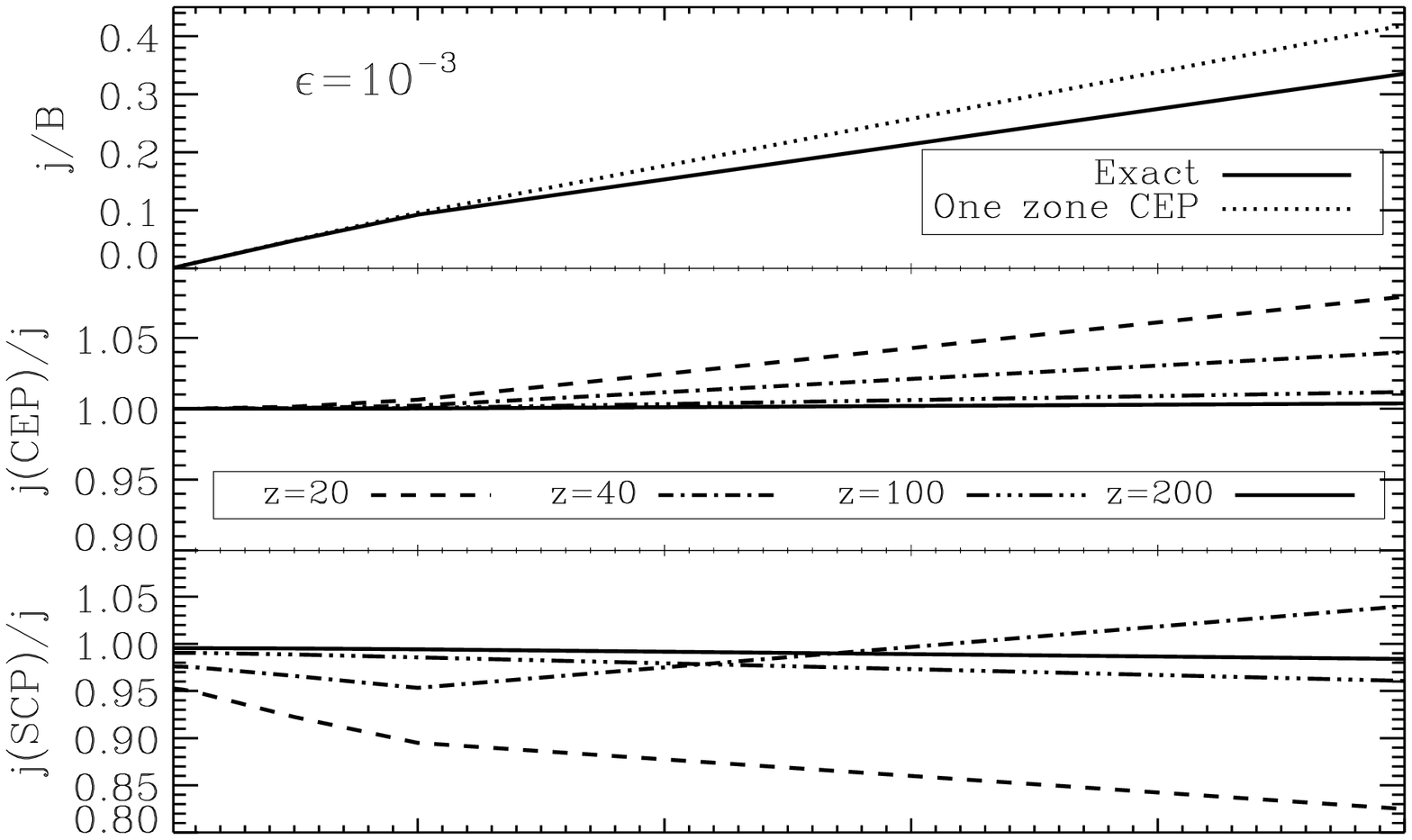}
 \includegraphics[width=\hsize,clip]{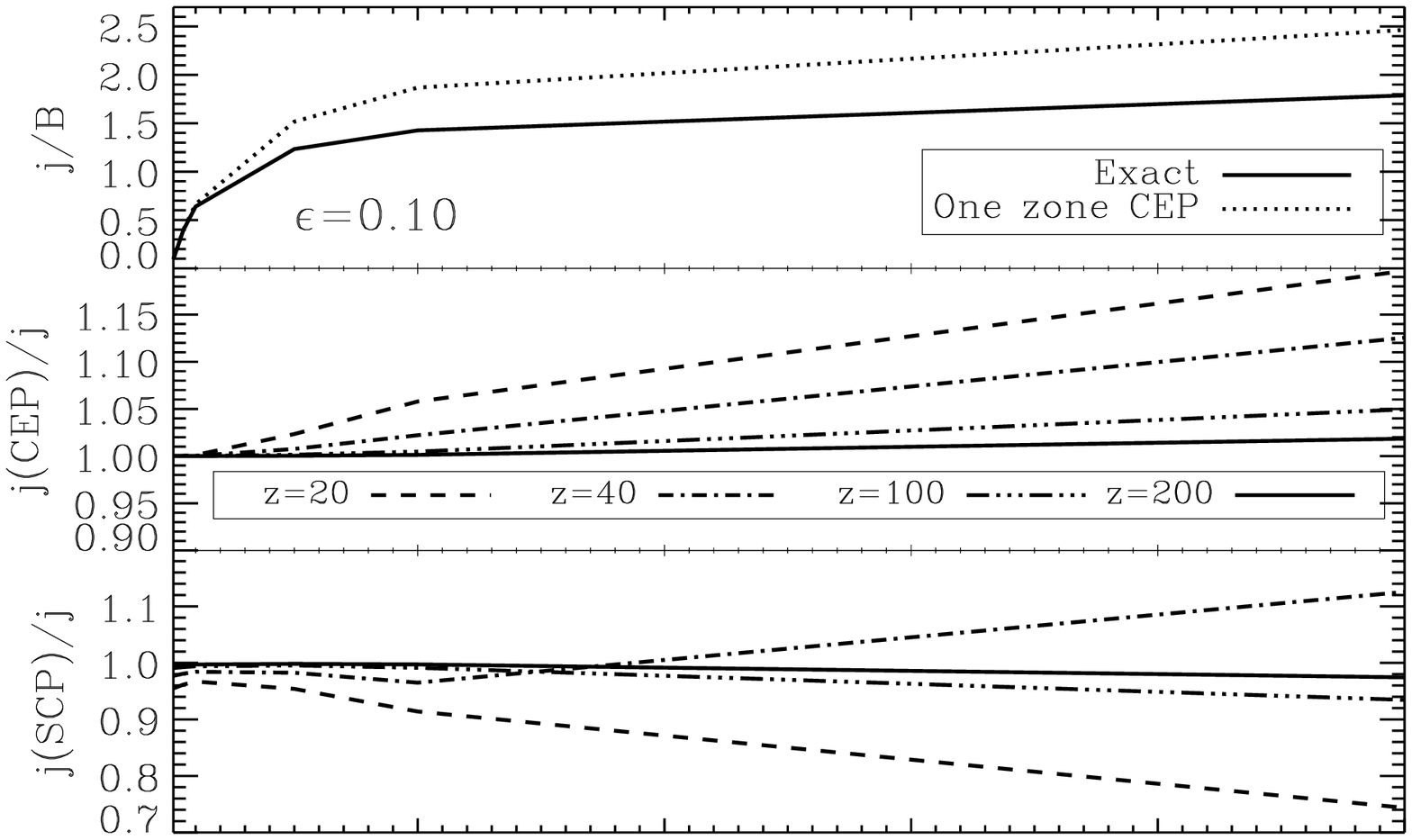}
 \includegraphics[width=\hsize,clip]{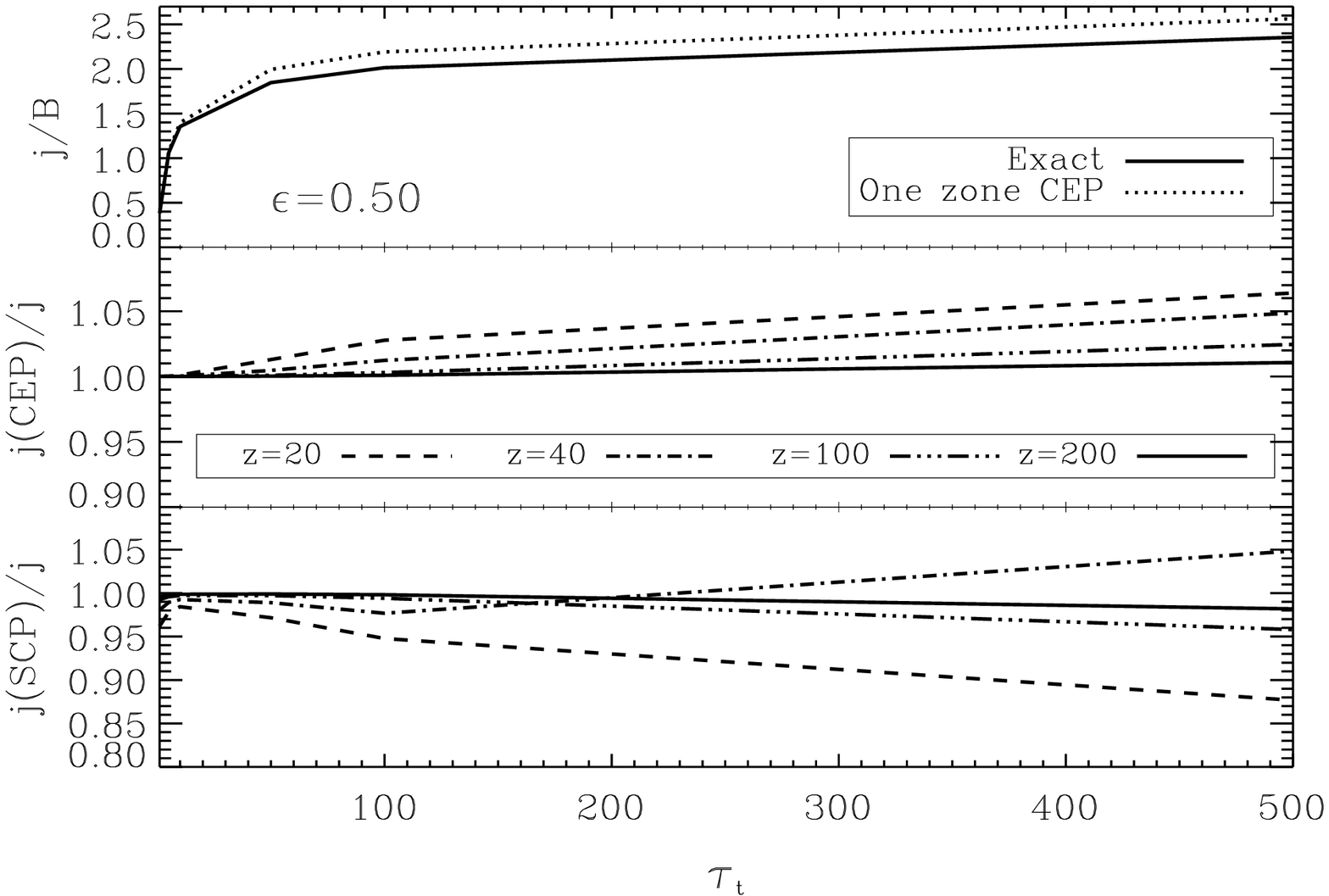}
\caption{The line cooling coefficient of a slab as a function of its overall
optical depth in two-level models with various $\epsilon$. The top panel of
each plot shows the exact solution and the result of a single-zone CEP
calculation. The two other panels show the convergence to the exact solution
with the number of zones $z$ for the CEP and SCP methods.}
 \label{fig:slab_j}
\end{figure}


\subsubsection{Slab emissivity}
\label{sec:slab_j}

The advantages offered by the CEP method are even more pronounced for slab
emission calculations. Figure \ref{fig:slab_j} shows a striking result: When
$\epsilon$ = \E{-5} (i.e., $\eta = \E5$), the CEP method produces with a single
zone the correct line cooling coefficient for all slabs with $\tot \le 500$!
This result is easy to understand. From the behavior of the function $\beta$ at
small and large $\tau$ (Capriotti 1965) it follows that $p(\tau) \sim 1$ when
$\tau < 1$ and that $p(\tau) \sim 1/\tau$ when $\tau \gg 1$, so that $p(\tau)
\ga 1/\tot$. Therefore, slabs with $\eta > \tot$ have $\eta p > 1$ everywhere.
From equations \ref{eq:S2} and \ref{eq:j} it follows that in this case
\eqarray{\label{eq:j_thin}
  \tot\ \ll \eta:&                                           \\
     S(\tau) &= \DS{B\over \eta p},         \hskip0.3in
    \j(\tot) = \DS\int_0^{\tot}\frac{1}{\eta}Bd\tau \nonumber
}
The expression for \j\ involves only input properties. That is, {\em the line
cooling coefficient can be calculated in this regime without even solving the
problem}. This result does not seem to have been recognized in the published
literature. When the physical conditions are constant, $\j = B\tot/\eta$; the
source is optically thick yet its emission increases linearly with optical
depth. The slab remains ``effectively-thin'' at large optical depths as long as
$\tot \ll \eta$ (i.e., $\epsilon\tot \ll 1$). And because the CEP method
employs discretized forms of these expressions, it reproduces the correct line
emission irrespective of the division into zones.

\begin{figure}
 \centering\leavevmode
 \includegraphics[width=\hsize,clip]{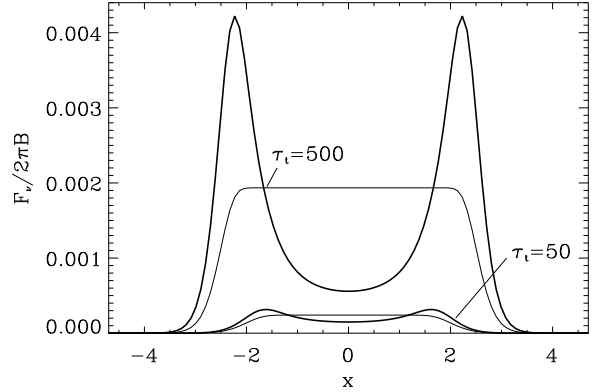}
\caption{Spectral shape of the flux emerging from slabs with $\epsilon$ =
\E{-5} and optical depths as marked. The thick lines are the result of the
exact solution, the thin lines of a  single zone CEP calculation.}
 \label{fig:profile}
\end{figure}

Since $\j(\tot)$ can be calculated without solving any equations, the
single-zone calculation produces the correct emission even though it does not
reproduce the correct population distribution --- as is evident from both eq.\
\ref{eq:j_thin} and figure \ref{fig:slab_S_epsE-5}, the source function varies
in the slab while the one-zone result is constant. Still, this constant value
is just the right average to reproduce correctly the slab luminosity. Another
perspective on this result is provided by the spectral shape of the emergent
radiation, shown in figure \ref{fig:profile}.  The exact solution properly
displays a self-absorption dip around line center (see, e.g., Avrett \& Hummer
1965). The single-zone calculation is incapable of producing this feature, but
its flat-top shape does enclose the same area, reproducing the correct line
luminosity. The simple one-zone calculation properly reproduces the overall
number of photons emitted in the line, though not the frequencies where they
emerge.

When the problem is formulated in terms of $\tau$, eq.\ \ref{eq:j_thin} gives
the line emission directly from the input properties. When the problem is
formulated instead in terms of densities and distances, eq.\ \ref{eq:j_thin}
implies that\footnote{This result was noticed in the limit in which $n_2 \ll
n_1$ by D.\ Neufeld in benchmark testing of radiative transfer codes, posted at
http://www.mpifr-bonn.mpg.de/staff/fvandertak/H2O/radxfrtest.pdf. Note that the
line cooling always obeys
\[
 \Lambda = E_{21}\int (C_{12}N_1 - C_{21}N_2)\, d\ell
\]
\[
  \phantom{\Lambda} = E_{21}\int g_1 C_{12}\, (n_1 - n_2e^{E_{21}/kT})\, d\ell
\]
as is evident from eqs.\ \ref{eq:j}, \ref{eq:S} and \ref{eq:R21}.}

\eq{
    \Lambda = E_{21}\int g_1 C_{12}\, (n_1 - n_2)\, d\ell.
}
Although the condition $\eta p > 1$ ensures that $n_2 \ll n_1$ when $E_{21} >
kT$, $n_2$ need not be negligible when $E_{21} < kT$. Therefore, the solution
must be executed in this case to determine the population distribution and the
actual value of $\tot$. Since the single-zone calculation does not produce the
correct population distribution, its result for the overall optical depth can
be wrong. To ensure the correct assignment of $\tot$ to the prescribed input,
the problem must be solved properly, including the division into zones.

\begin{figure}
 \centering\leavevmode
 \includegraphics[width=\hsize,clip]{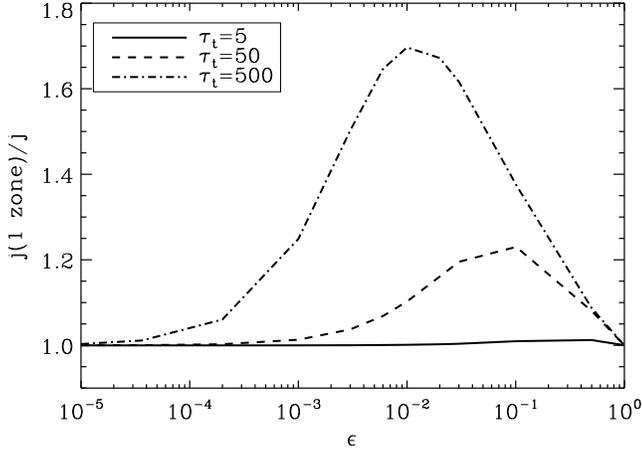}
\caption{Deviation of the one-zone CEP result from the exact value of the line
cooling coefficient as a function of $\epsilon$ in two-level models with
various slab thicknesses.}
 \label{fig:j_eps}
\end{figure}

When $\epsilon$ increases, the slab ceases to be ``effectively thin" and the
line luminosity begins to deviate from the one-zone CEP result, as is evident
from figure \ref{fig:slab_j}. Eventually, line thermalization sets in with
further increase in $\epsilon$, and the single-zone result again becomes
adequate. This behavior is further illustrated in figure \ref{fig:j_eps}. At a
fixed $\tot$, the deviation from the single-zone CEP calculation reaches a
maximum when $\epsilon \sim 5/\tot$. When $\tot$ = 50 the maximal deviation is
\about\ 20\% at $\epsilon$ \about\ 0.1, when $\tot$ = 500 it is \about\ 70\% at
$\epsilon$ \about\ 0.01. Varying $\epsilon$ away from that peak in either
direction, the one-zone CEP calculation gives a progressively better
approximation.

With \Ncr\ = 4\x\E3\ cm$^{-3}$ at $T$ = 75 K, the 158 \mic\ line of \ion{C}{II}
is in the regime $N \ga \Ncr$ ($\epsilon \ga 0.5$) in most cases of interest.
Therefore single-zone CEP calculations for this line are expected to produce
cooling rates accurate to better than \about\ 10\% under most circumstances.



\section{Multi-Level Systems}
\label{sec:multi}

\def\t(#1){\tau^{#1}_{ul}}
\def\a(#1){\alpha^{#1}_{ul}}

Consider $L$ energy levels. A trivial change from the two-level case is the
addition of some bookkeeping indices. We designate level numbers with
subscripts and zone numbers with superscripts. In zone $i$, the population per
sub-state of level $k$ is $n_k^i$ and the overall population is
\eq{\label{eq:Norm}
    n^i = \sum_{k = 1}^L g_k n_k^i
}
where $g_k$ is the level degeneracy. Unlike the two-level system, locations in
the slab cannot be specified by optical depth anymore because each transition
has a different optical depth, which can be determined only after the unknown
$n_k^i$ are calculated. Instead, the partition into zones is done in terms of
the geometrical distance from one surface. Denote by $\ell^i$ the width of the
$i$-th zone, then its optical thickness in the transition between lower level
$l$ and upper level $u$ is
\eq{
    \tau_{ul}^{i, i - 1} = {1\over4\pi\DnuD}g_uB_{ul}E_{ul}
        \left(n_l^i - n_u^i\right)\ell^i,
}
where $E_{ul}$ is the energy separation between levels $u$ and $l$. The
equivalent of equation \ref{eq:tij} is then
\eq{
    \tau_{ul}^{i,j} = \sum_{k = j + 1}^i\tau_{ul}^{k, k - 1}
}
when $i > j$. In complete analogy with equations \ref{eq:rate}, \ref{eq:pi} and
\ref{eq:M}, the level population equations are
\eqarray{\label{eq:final}
    {dn_k^i\over dt} = -\sum_{l = 1}^{k - 1} A_{kl}p_{kl}^i n_k^i +
      C_{kl}^i\left(n_k^i - n_l^i e^{-E_{kl}/kT^i}\right) \qquad      \\
  \hskip0.2in   +\sum^{L}_{u = k + 1} {g_u\over g_k}
       \left[A_{uk}p_{uk}^i n_u^i +
      C_{uk}^i\left(n_u^i - n_k^i e^{-E_{uk}/kT^i}\right)\right]   \nonumber
}
Here
\eq{\label{eq:p_ul}
     p^i_{ul} = \beta^i_{ul} + {1\over\t(i,i-1)}
     \sum_{\stackrel{j = 1}{j \ne i}}^z
     {n_u^j\over n_u^i}\,{n_l^i - n_u^i\over n_l^j - n_u^j}  M_{ul}^{ij}
}
where
\eq{
    M_{ul}^{ij} = -\frac12(\a(i,j) - \a(i-1,j) - \a(i,j-1) + \a(i-1,j-1))
}
and where $\beta_{ul}^i = \beta(\t(i,i-1))$ and $\a(i,j) =
\t(i,j)\beta(\t(i,j))$. This provides a set of $L - 1$ independent equations
for the $L$ unknown populations in each zone, $n_k^i$. Equation \ref{eq:Norm}
for the overall density in the zone closes the system. The overall system of
non-linear algebraic equations for the level populations in all zones is
readily solved with the Newton method.

It is convenient to switch to the scaled quantities $n_k^i/n^i$ as the unknown
variables and introduce the overall column density
\eq{
    \N = \sum_{i = 1}^z  n^i\ell^i
}
Neither densities nor physical dimensions need then be specified since only \N\
enters as an independent variable; the zone partition is done in terms of \N\
rather than $\ell$. The problem is fully specified by three input parameters:
density $N$ of collision partners and temperature $T$, which together determine
the collision terms, and \N\ (in fact, \N/\DnuD), which sets the scale for all
optical depths.


\begin{figure}
 \centering\leavevmode
 \includegraphics[width=\hsize,clip]{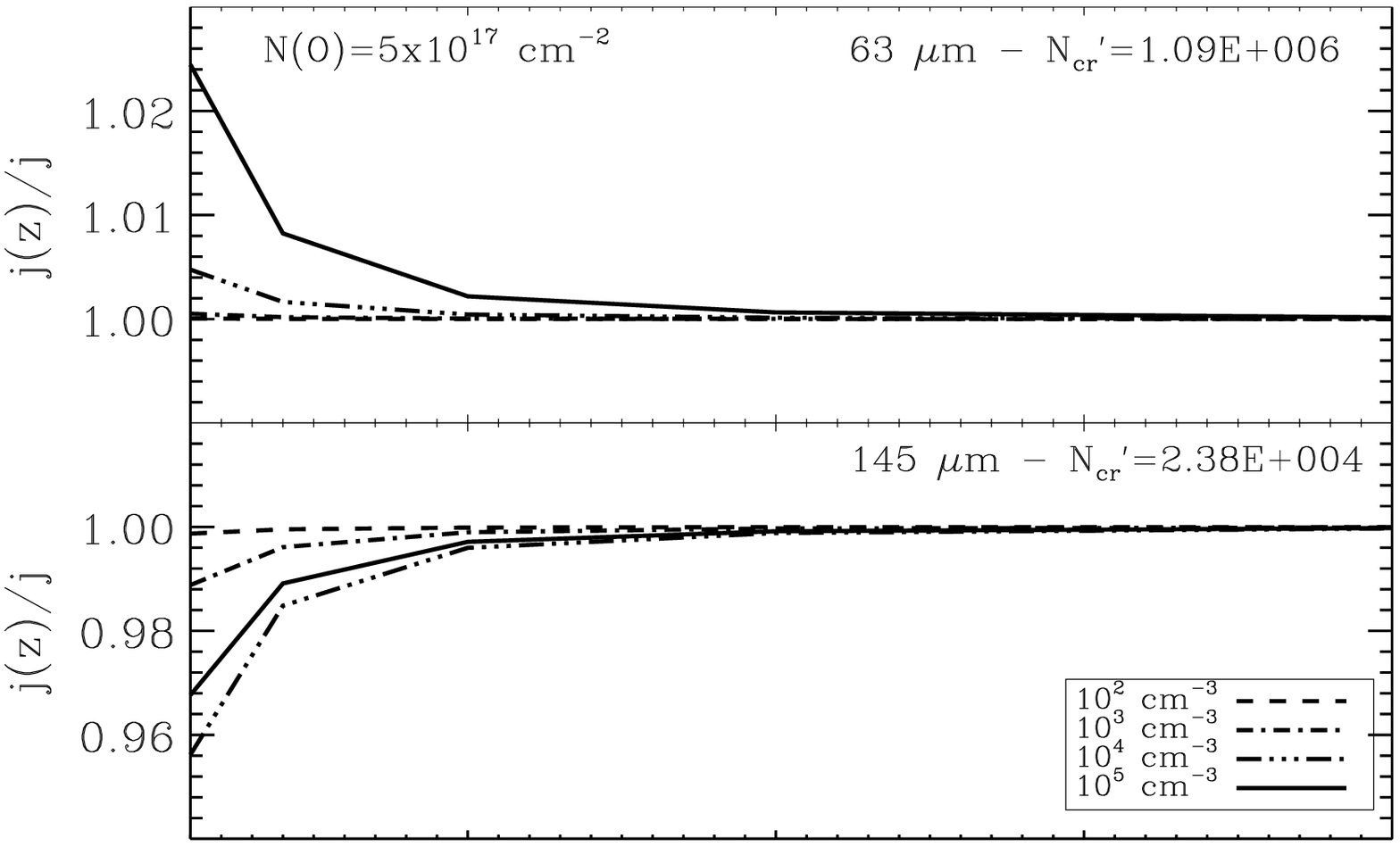}
 \includegraphics[width=\hsize,clip]{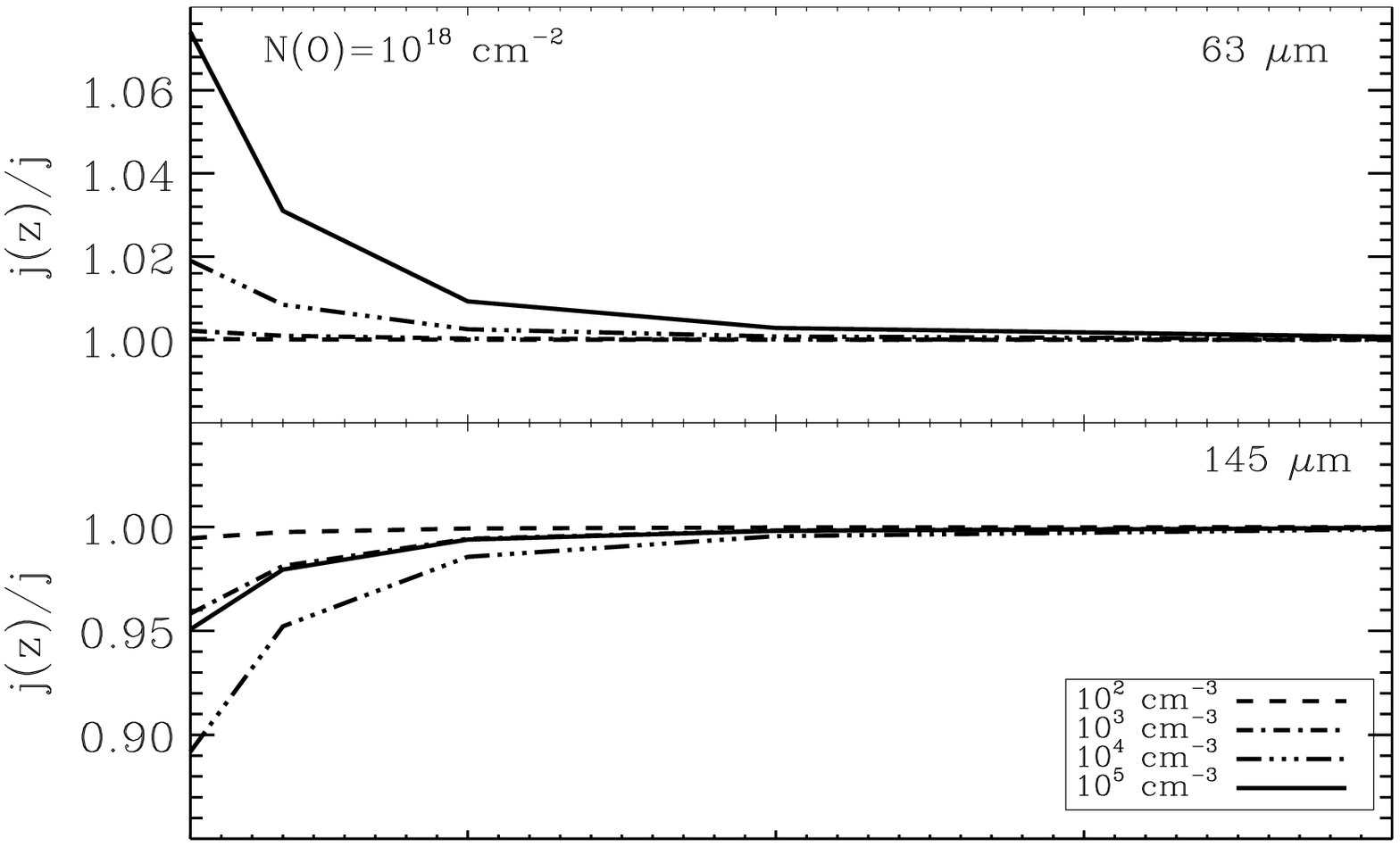}
 \includegraphics[width=\hsize,clip]{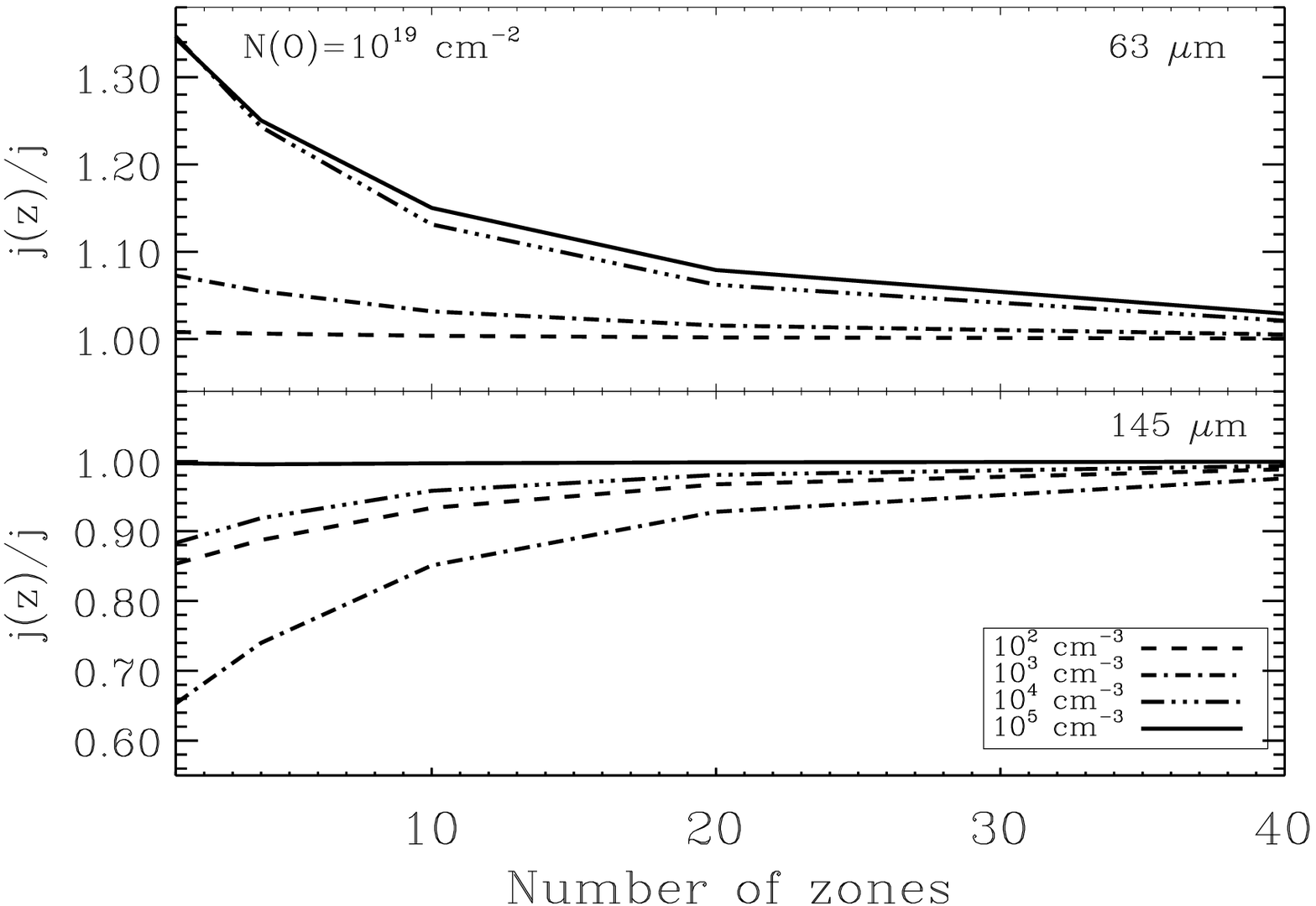}
\caption{Variation of the emission in the $^3$P cooling lines of \ion{O}{I}
with the number of zones in CEP calculations. Shown are the ratios to the exact
solutions for slabs with $T$ = 100 K, various H densities, as marked, and three
representative oxygen column densities. At \N(O) $\le$ \E{17} cm$^{-2}$, the
single-zone calculation produces the exact result for both lines. The critical
density for each line is listed in the top figure.}
 \label{fig:OI_zones}
\end{figure}

%
%
%
%



\begin{figure*}
\begin{minipage}{\textwidth}
\centering \leavevmode
 \includegraphics[bb=0 45 498 340,width=0.48\hsize,clip]{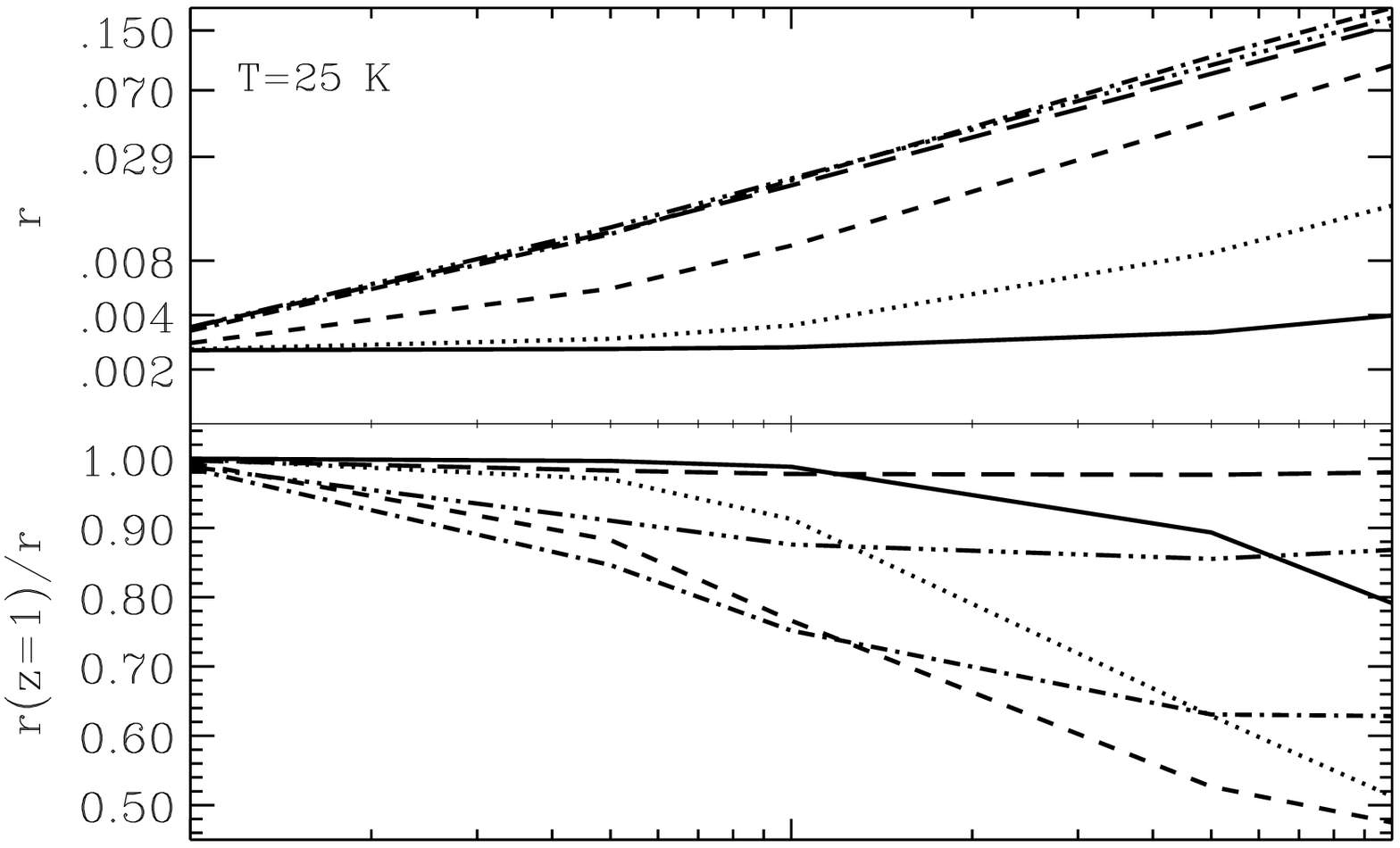}\hfil
 \includegraphics[bb=0 45 498 340,width=0.48\hsize,clip]{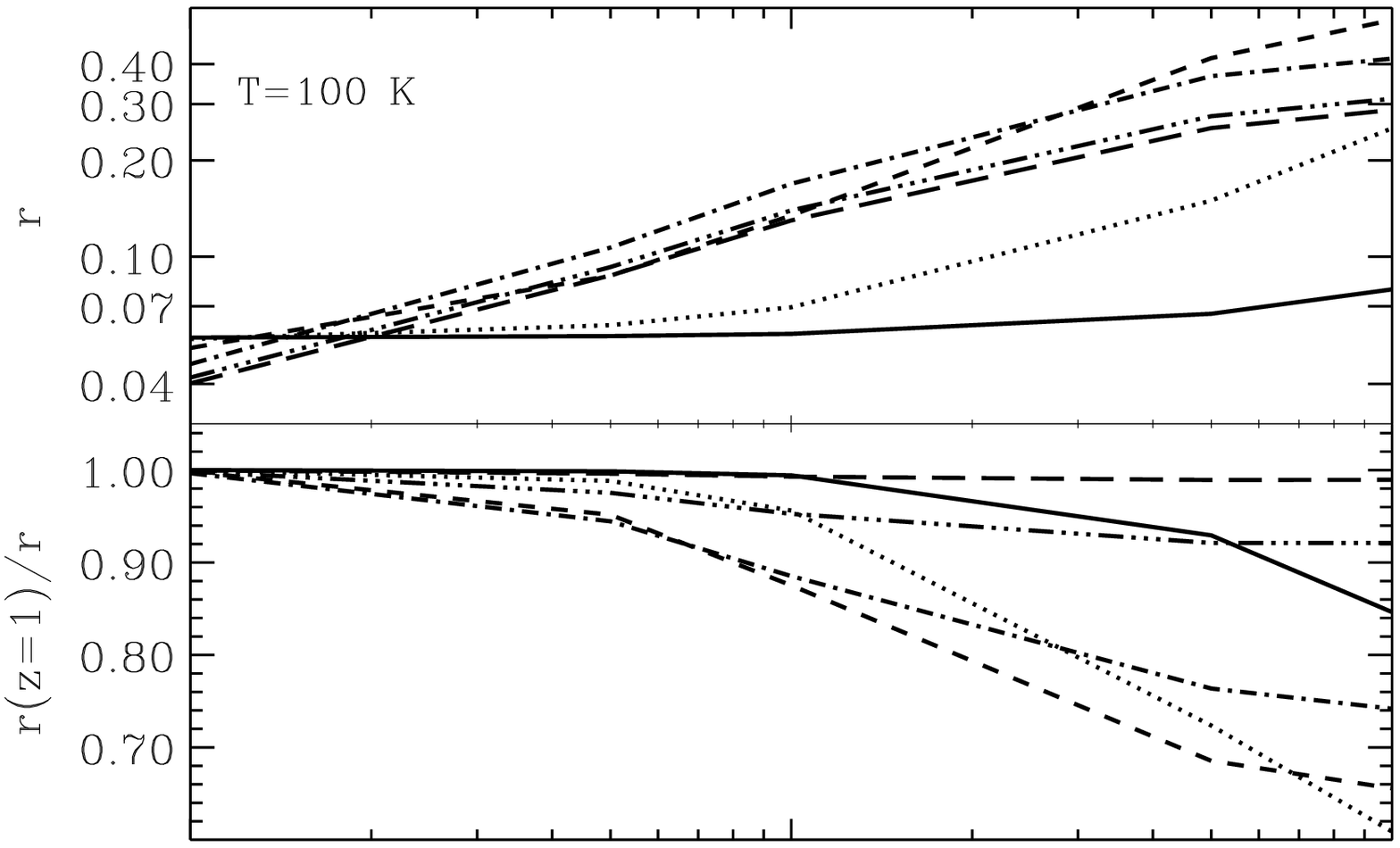}\\
 \includegraphics[bb=0 45 498 340,width=0.48\hsize,clip]{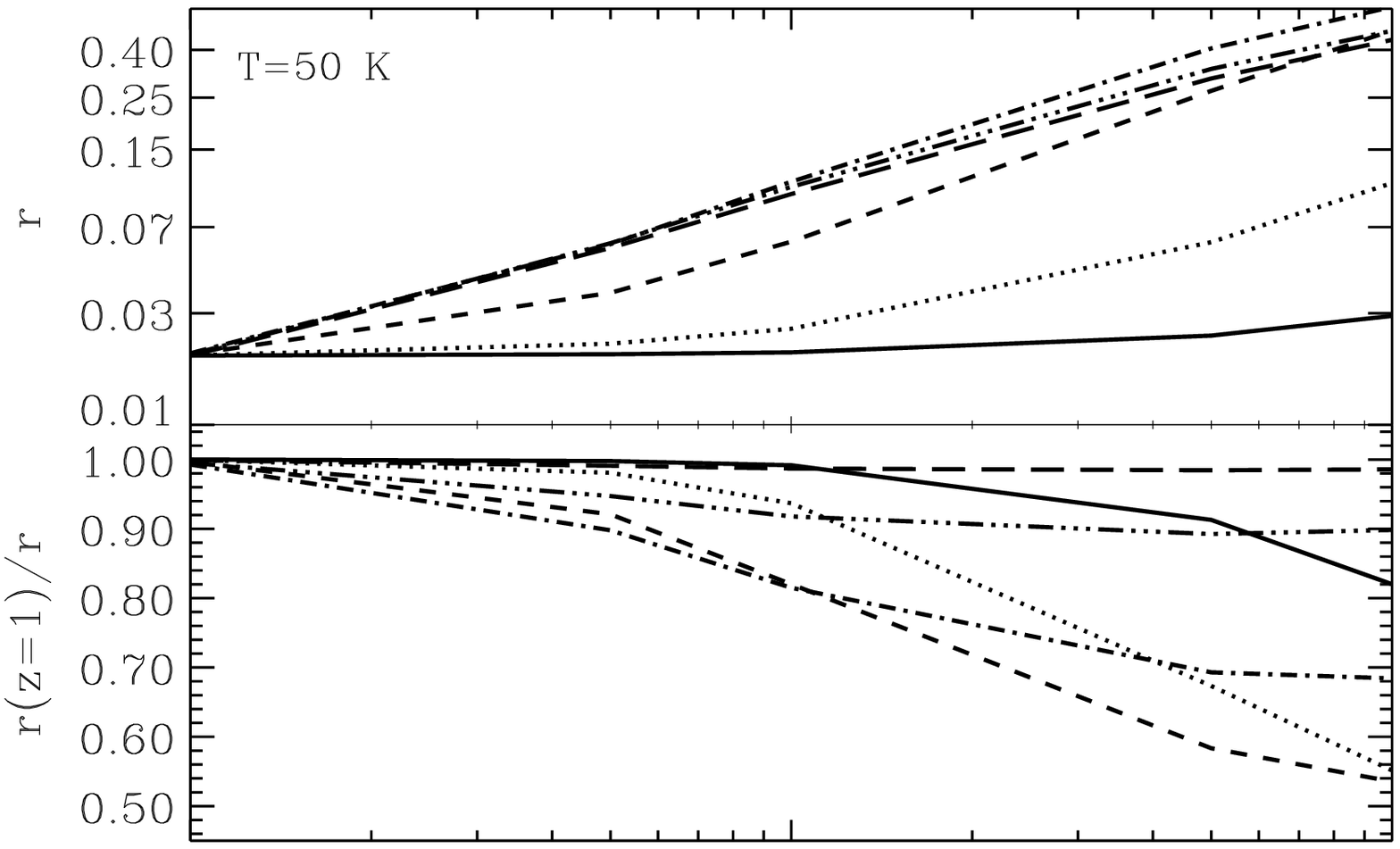}\hfil
 \includegraphics[bb=0 45 498 340,width=0.48\hsize,clip]{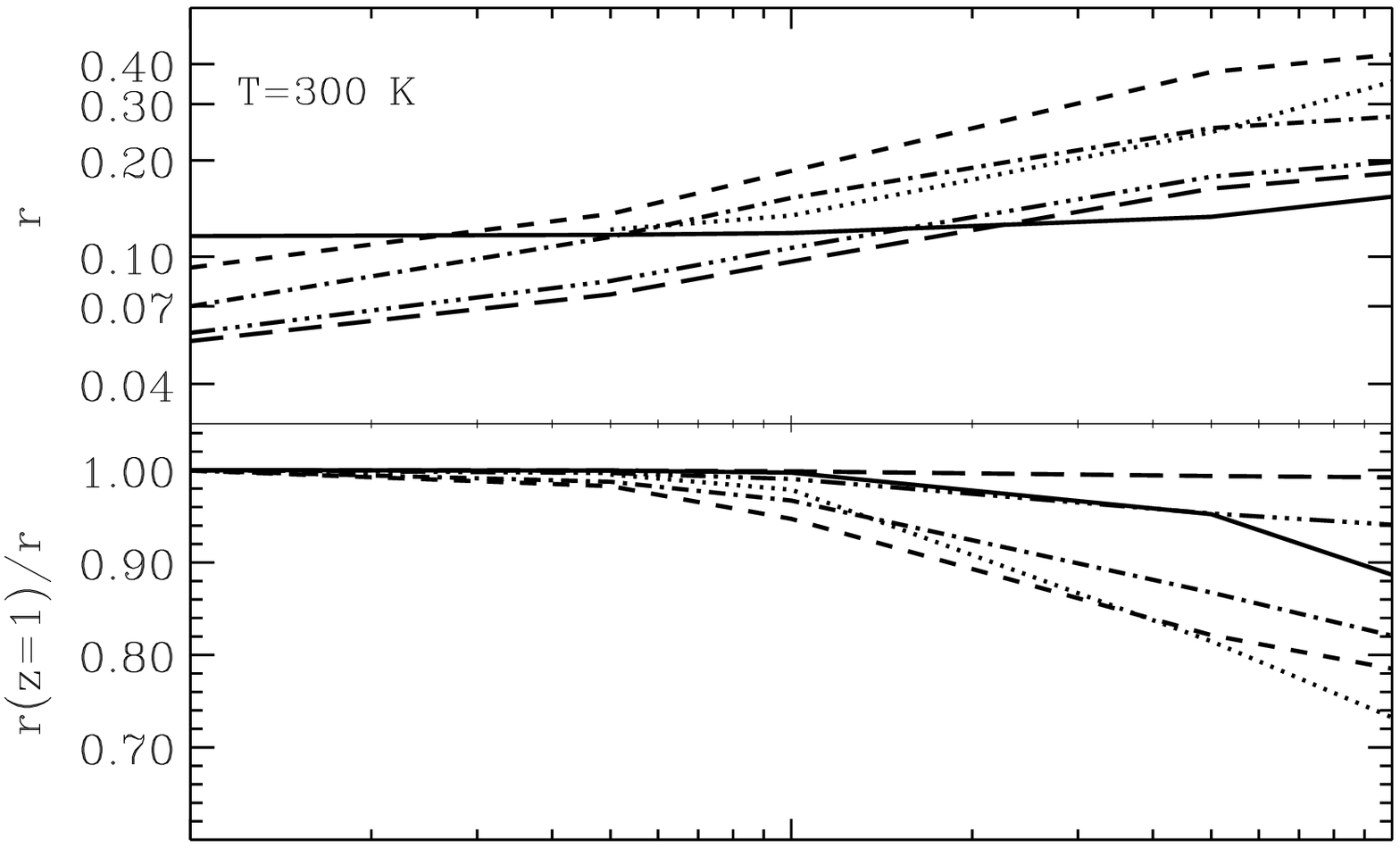}\\
 \includegraphics[bb=0  0 498 340,width=0.48\hsize,clip]{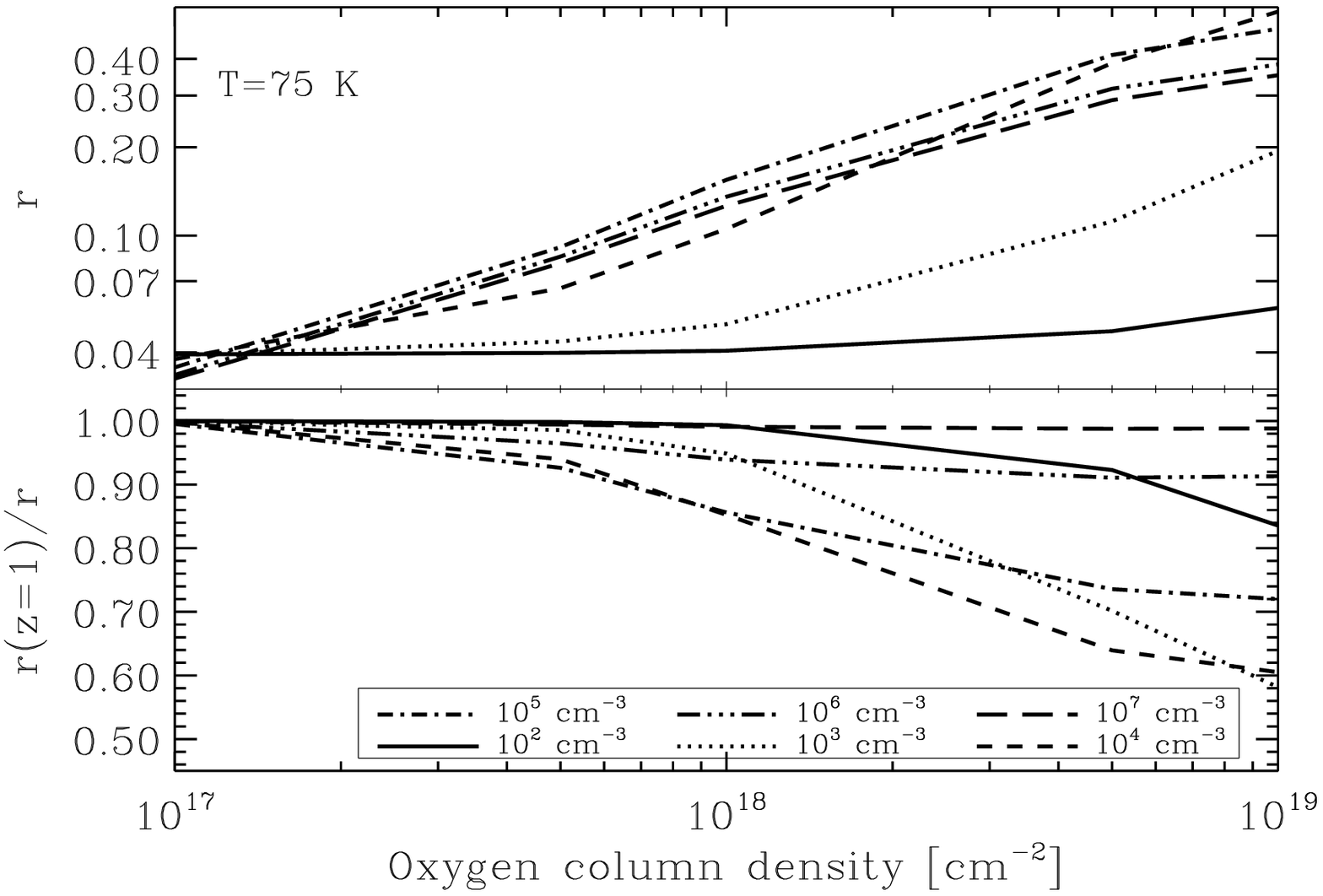}\hfil
 \includegraphics[bb=0  0 498 340,width=0.48\hsize,clip]{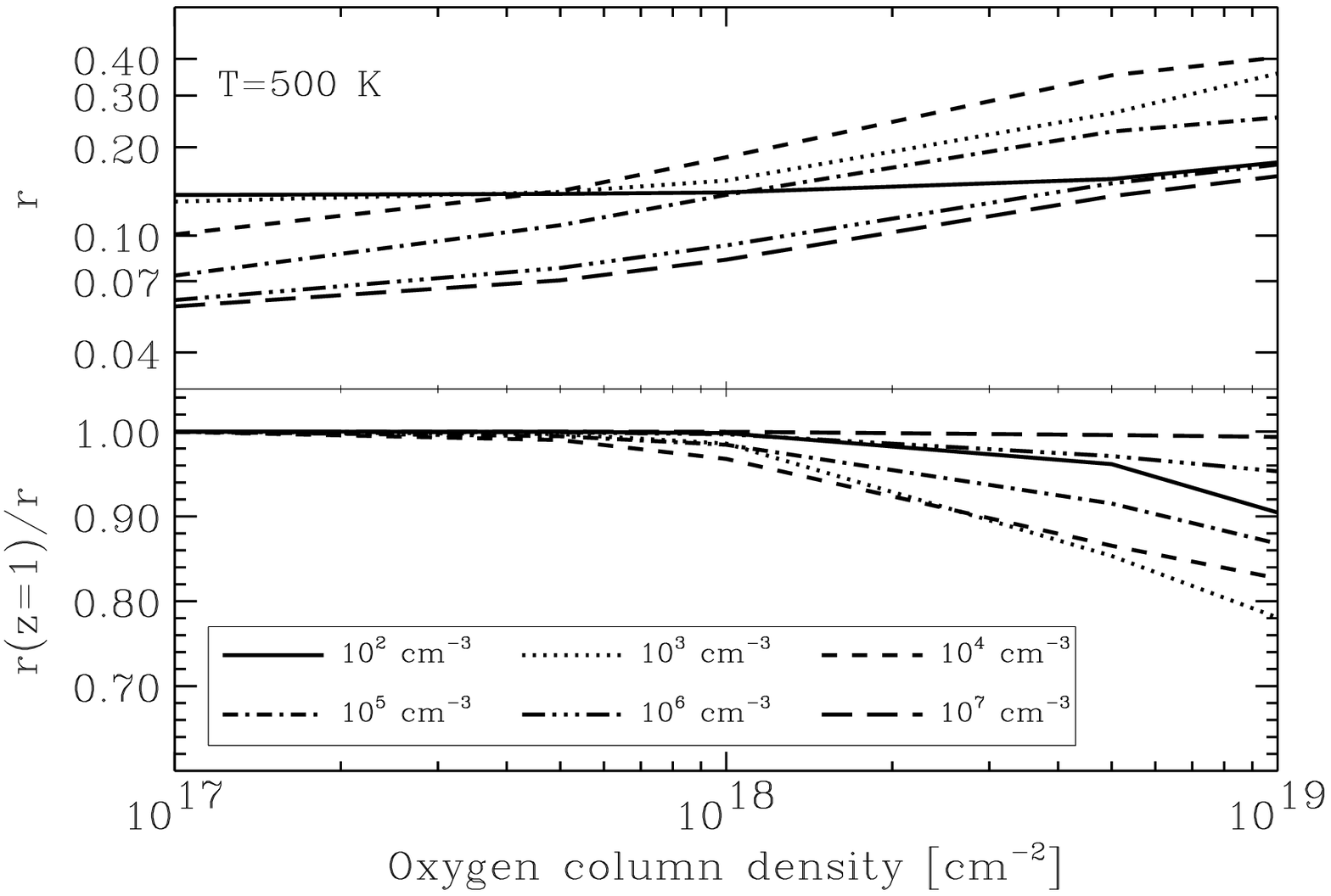}
\caption{The ratio $r = \j(145\mic)/\j(63\mic)$ of the \ion{O}{I} cooling lines
as a function of oxygen column density for various temperatures and H
densities, as marked. The top panel in each case shows the exact ratio $r$, the
bottom panel the ratio of the results of single-zone calculations to the exact
ones.} \label{fig:OI_ratios}
\end{minipage}
\end{figure*}

\subsection{Example --- the O\,{\small I} cooling lines}

Together with the \ion{C}{II} 158 \mic, the $^3$P lines of \ion{O}{I} at 63
\mic\ and 145 \mic\ dominate the gas cooling of warm PDR. Ratios and peak
intensities of these lines are used to measure the gas density and temperature
(Tielens \& Hollenbach 1985). In \S\ref{sec:slab_j} we found that simple escape
probability calculations do reproduce the proper \ion{C}{II} 158 \mic\
emission. We now examine the behavior of \ion{O}{I} lines through an exact CEP
calculation of the three levels of the $^3$P system. We solve for slabs with
constant physical conditions, specified by temperature, hydrogen density and
oxygen column density \N(O).

Figure \ref{fig:OI_zones} shows the effect of varying the number of zones on
the cooling lines emission at $T$ = 100 K; the results are similar for $T$ =
300 K and 500 K.\footnote{It is interesting to note that the 145 \mic\
transition undergoes population inversion in the optically thin regime at low
densities for temperatures above 300 K. The reason is that the radiative
lifetime of its upper level is more than five times longer than for its lower
level.} Single-zone CEP calculations produce the exact result at $\N(O) \le
\E{17}$ cm$^{-2}$, but deviate from it at larger column densities by amounts
that increase with \N(O). The deviation is different for each line, reflecting
their different critical densities, which are listed in the figure, and optical
depths; for reference, at \N(O) = \E{19} cm$^{-2}$ $\tau(63\mic)$ \about\ 100
while $\tau(145\mic)$ varies from \about 0.6 to \about 2, depending on the
density. Because of the different trends displayed by the emission in the two
cooling lines, the one-zone calculation generally misses the exact result for
their ratio $r = \j(145\mic)/\j(63\mic)$. Figure \ref{fig:OI_ratios} shows the
variation of $r$ with column density and the deviations of the results of a
one-zone calculation from the exact values. The single-zone results differ from
the exact values by amounts that vary with temperature, density and oxygen
column density. The deviations are generally largest at hydrogen densities
around \E3\ -- \E4\ cm$^{-3}$. As is evident from this figure, the variation
range of $r$ is comparable to the error that can be introduced by its
calculation with a single-zone. A reliance on such calculations can lead to
erroneous conclusions regarding the physical conditions in a source. Indeed,
from the observed ratio of the two [OI] lines and escape probability
calculations, Caux et al (1999) deduced a mean gas temperature of 26$\pm$0.5 K,
an H$_2$ density $\ge$ 3\x\E4 cm$^{-3}$ and an [OI] column density $\ge$
5\x\E{19} cm$^{-2}$. However, as is evident from figure \ref{fig:OI_ratios},
the large values which they observed ($r \sim 0.4$) can be reached at lower
columns for a wide range of somewhat higher temperatures.

One-zone CEP calculations of the \ion{O}{I} cooling lines are not as reliable
as they are for the \ion{C}{II} 158 \mic. However, it does not take too many
zones for the CEP method to achieve adequate accuracy.  As is evident from
figure \ref{fig:OI_zones}, 20 zones suffice for accomplishing better than 10\%
accuracy, and the exact solution is reached to within 1\% with 40 zones in all
cases.

\section{Discussion}

The test cases presented here show that our new method not only provides an
exact solution of the line transfer problem, it also outperforms the leading
ALI solver by a margin even larger than that among different implementations of
the ALI technique. Two fundamental properties give the CEP method intrinsic
advantages. The first is the calculation of \Jbar, the only radiative quantity
required for solving the level populations. In the standard approach,
$I_\nu(\mu)$ is determined from a solution of eq.\ \ref{eq:rad_tran} and \Jbar\
is calculated from eq.\ \ref{eq:Jbar} in angular and frequency integrations
that involve a-priori unknown dependencies on these two variables. Determining
the dependence of $I_\nu(\mu)$ on $\nu$ and $\mu$ is a major task for the
solution of the radiative transfer equation. Deviations of the computed
$I_\nu(\mu)$ from its exact angular shape and frequency profile contribute to
the error in the computed \Jbar\ in each iteration. In contrast, the CEP method
determines \Jbar\ from the integration in eq.\ \ref{eq:p2} that involves known
dependencies on both $\nu$ and $\mu$; the dependence on these variables enters
only from the optical depth $\tau_\nu(\mu)$, and it is a-priori known from the
input to the problem. The angular and frequency integrations are exact in the
CEP method; the fact that the dependence of $I_\nu(\mu)$ on $\nu$ and $\mu$ is
unknown is altogether irrelevant.

The other intrinsic strength of the CEP method is that it involves only level
populations and thus takes full advantage of thermalization wherever that sets
in. In contrast, the ALI technique must repeatedly solve the radiative transfer
equation in the entire source, even in thermalized regions, to determine the
radiation field everywhere.

\subsection{Technicalities}
\label{sec:tech}

The great efficiency of the Newton method in solving non-linear equations is
another advantage of the CEP method. The prerequisite for a successful solution
is a reasonable initial guess. An efficient strategy for working
implementations of CEP is to start from the actual populations of a previous
solution for similar physical conditions. A particularly useful approach is to
start from the optically thin limit in which $p$ = 1 everywhere and solve the
level populations from the corresponding linear equations. The column density
\N\ is now increased in small steps until the desired value is reached, using
in each step the populations from the previous one as the initial distribution.
This technique can also work in the opposite direction
--- start from thermal equilibrium populations and a very large \N, and
decrease \N\ in small steps. An added advantage of this approach is that each
step also provides information about the number of zones required for CEP
convergence.

The Newton method requires inversion of the Jacobian matrix, and the number of
operations in this process increases as the third power of the matrix
dimension. Although this rapid rise did not have a serious effect in the
examples presented here, it could degrade the performance in cases of very
large numbers of levels and zones. Matrix inversion is avoided in the iterative
scheme designed by van der Vorst (1992) for solution of the linear system of
the Newton method. In this scheme, geared toward sparse Jacobian matrices, only
the non-zero matrix elements are stored and used. We have experimented with
this method and found it to be quite useful for the CEP technique. It is
particularly suitable for multi-level problems because they tend to produce
sparse matrices, as each level generally couples to only a limited number of
other levels. Other alternatives are to use quasi-Newton schemes (e.g.,
Broyden's method) like those employed by Koesterke et al.\ (1992), or evolve
the set of differential equations \ref{eq:final} until reaching steady state.

The efficiency of CEP computations can be further enhanced with better grid
design. Our solutions of the semi-infinite atmosphere employed grids with equal
logarithmic spacing in $\tau$ over ten orders of magnitude, resulting in
extremely thick zones deep inside the atmosphere. For example, even with $z$ =
600, the zone thickness was $\tau = 1.7\x\E5$ in the $\E6 \le \tau \le \E7$
region. These extremely thick zones do not pose any difficulties to CEP
computations because they occur in regions where the populations are
thermalized. Indeed, the zones could be even thicker in a much larger fraction
of the source without compromising accuracy. It should thus be possible to
achieve the same accuracy with fewer zones by concentrating them in the regions
were the populations deviate from thermal equilibrium. Since the number of
zones is the single most important factor in determining CEP runtime, a more
sophisticated grid construction will make the method even more efficient. We
intend to investigate the implementation of adaptive gridding algorithms in
future work.

An additional increase in efficiency can be easily gained in practical
applications that do not require the extreme precision we imposed in this
comparative study. Here the functions $\alpha$ and $\beta$ were calculated
using the integral definition in eq.\ \ref{eq:alpha}, repeatedly performing
highly accurate quadrature. Instead, one could employ the approximate series
expansion derived by Capriotti (1965) for the function $\beta$. We have
verified that this rapidly convergent series is always within 3\% of the exact
result.\footnote{In Capriotti (1965), the small- and large-$\tau$ portions of
the expansion were joined at line-center optical depth $\tau_0$ = 5. They
should be joined instead at $\tau_0$ = 3.41 for a smooth transition.} Another
option is to calculate once a finely spaced table of the $\alpha$ and $\beta$
functions and interpolate between its elements with an efficient algorithm.

\subsection{CEP and ALI}
\label{sec:CEP-ALI}

The CEP method is suitable for solution also with the ALI approach. Starting
from eq.\ \ref{eq:Jbar} in the operator form $\Jbar = \Lambda S$, the ALI
technique is based on the operator splitting $\Lambda = \Lambda^* +
(\Lambda-\Lambda^*)$, where $\Lambda^*$ is an approximation to the $\Lambda$
operator. The mean intensity is obtained from the approximate expression $\Jbar
= \Lambda^*S + (\Lambda-\Lambda^*) S^{\rm prev}$ that involves the source
function from the current and previous iterations. This approximation becomes
exact upon convergence, when $S = S^{\rm prev}$. As already noted, it has been
shown that the optimal choice for $\Lambda^*$ is the diagonal of $\Lambda$.

Equation \ref{eq:p1} gives $\Jbar = (1 - p)S$, the $\Lambda$-operator form of
the CEP method. From eq. \ref{eq:pi}, the matrix elements of the CEP
$\Lambda$-operator are simply
\eq{
   \Lambda_{ij} = (1-\beta^i) \delta_{ij}
      - {M^{ij}\over\tau^{i,i-1}}(1 - \delta_{ij})
}
where $\delta_{ij}$ is the Kronecker delta. Thanks to the known dependence on
$\nu$ and $\mu$ in the CEP approach, this expression allows the usage of
approximate operators $\Lambda^*$ of increasing complexity without any
additional computational effort in the calculation of the matrix elements. In
standard ALI techniques $\Lambda^*$ is the diagonal of $\Lambda$, i.e.,
$\Lambda^*_{ij} = (1 - \beta^i)\delta_{ij}$. We have implemented this choice in
an ALI solution of eq.\ \ref{eq:S2a} and performed the two-level model
calculations presented in this paper also with this technique. In all cases,
the solution converged to the exact same results as the algebraic CEP equations
\ref{eq:S2CEP}. Runtime for this ALI implementation of the CEP method was on
par with the SCP method up to 200 zones, but fell behind at larger
$z$.\footnote{We implemented the Ng (1974) acceleration technique to improve
the convergence rate of the ALI method in both SCP and CEP.} Given that in the
CEP ALI implementation we have adopted the standard choice for $\Lambda^*$, the
optimal choice in the CEP approach could well be different, improving the
performance.

It is also important to point out that the CEP method could, in principle, be
implemented in the framework of the Gauss-Seidel and Successive Overrelaxation
iterative methods. These methods were first applied in radiative transfer
problems by Trujillo Bueno \& Fabiani Bendicho (1995) using SCP as the formal
solver. They can lead to an order of magnitude improvement in the number of
iterations used to reach convergence, with a time per iteration that is
virtually the same as the method based on the Jacobi iteration. Also of
interest is the possibility of implementing the CEP method in the linear
(Steiner 1991) or the non-linear (Fabiani Bendicho, Trujillo Bueno \& Auer
1997) multi-grid methods. All of these issues will be addressed in future
investigations.

\subsection{Extensions}

All the examples presented in this paper involved constant physical conditions.
Variable conditions are handled by simply starting with zones that have
constant physical conditions and proceeding to refine those divisions as
required by the CEP solution accuracy. Equation \ref{eq:final} for the level
populations already incorporates the handling of variable conditions by
allowing the temperature and collision rates to vary between the zones.

For simplicity, our method was introduced in the context of a quiescent slab
with the Doppler line profile. None of these simplifications represents an
inherent limitation of the CEP method. The formal expressions do not specify
the shape of $\Phi(x)$, and other line profiles can be implemented just the
same. Extension from the slab to other geometries, although straightforward,
requires some more work. Thanks to the planar symmetry, the angular variation
of optical depth in a slab is simply $\tau(\mu) = \tau(\mu = 1)/\mu$,
independent of either position or density profile. This symmetry does not carry
to any other geometry; even in the case of spherical symmetry, the angular
variation of $\tau$ cannot be calculated at any point other than the center
without specifying the density profile. However, generalizing the basic CEP
relation eq.\ \ref{eq:p2} to handle any geometry is straightforward, and the
fundamental advantage of integration over known frequency and angular
variations remains intact. Finally, recalling that large velocity gradients was
the context in which the escape probability approximation was originally
introduced by Sobolev, the CEP method is well suited for exact handling of this
case too.

The escape probability approach has been used in a number of simplified
calculations of complex problems. These include: overlapping of spectral lines
(so called ``line fluorescence"), important for various ionic transitions
(e.g., Bowen lines) in photoionized regions and OH lines in molecular clouds
(Guilloteau, Lucas \& Omont 1981, Elitzur \& Netzer 1985, Lockett \& Elitzur
1989); the effect of line overlap with underlying continuum (Netzer, Elitzur \&
Ferland 1985); and photoionization (Elitzur 1984). Importing these applications
into the CEP framework is straightforward. Finally, another extension is the
application of the CEP method to the self-consistent solution of the radiative
transfer equations for polarized radiation and of the statistical equilibrium
equations for the density matrix, the so-called non-LTE problem of the
2$^\textrm{nd}$ kind (see, e.g., Landi Degl'Innocenti 2003). We plan to provide
these extensions in future publications.

\subsection{Conclusions}

While our new method outperforms the current leading techniques, its greatest
advantage is its simplicity and ease of implementation. The CEP method employs
a set of algebraic equations (eq.\ \ref{eq:final}) that are already
incorporated in numerous widely used codes based on the escape probability
approximation. All that is required for an exact solution with these existing
codes is to augment the escape probability with the zone-coupling sum on the
right-hand-side of eq.\ \ref{eq:p_ul}. With this simple modification, the
multi-level line transfer problem is solved exactly.

\section*{Acknowledgments}

We thank Riccardo Cesaroni, Floris van der Tak and St\'ephane Guilloteau for
valuable remarks that helped us considerably in the preparation of this
manuscript. This research was supported by NSF, NASA, the European Commission
through the Solar Magnetism Network (contract HPRN-CT-2002-00313) and by the
Spanish Ministerio de Educaci\'on y Ciencia through project AYA2004-05792.

\let\Ref=\item
{}

\appendix

\section                   {External Radiation}
\label{sec:external}

The only effect of external radiation on the rate equations is to modify the
exchange rate $R_{ul}^i$ between levels $u$ and $l$ in the $i$-th zone (see
eq.\ \ref{eq:rate}) according to
\eq{
    R_{ul}^i \Rightarrow R_{ul}^i -B_{ul}\Ji(n_u^i - n_l^i),
}
where \Ji\ is the zone average (as in eq.\ \ref{eq:avg}) of the contribution of
the external radiation to the local $\Jbar$. When the external radiation
corresponds to the emission from dust which permeates the source, \Ji\ is
simply the angle-averaged intensity of the local dust emission in the $i$-th
zone. When the external radiation originates from outside the slab and has an
isotropic distribution with intensity \Ie\ ($= J_e$) in contact with the $\tau
= 0$ face, then
\eq{
  \Ji = \half J_e{1\over\t(i,i-1)}(\a(i,0) - \a(i-1,0)).
}
When the slab is illuminated by parallel rays with intensity \Ie\ ($= 4\pi
J_e$) entering at direction ($\mu_0,\phi_0$) to the $\tau = 0$ face then
\eq{
  \Ji = J_e{\mu_0\over\t(i,i-1)}
        \left[\gamma(\t(i)/\mu_0) - \gamma(\t(i-1)/\mu_0)\right]
}
where
\eq{
   \gamma(\tau) = \int_{-\infty}^{\infty}\left[1 - e^{-\tau\Phi(x)}\right]dx.
}

\label{lastpage}
\end{document}